\documentclass[aps,prd,preprint,tightenlines,superscriptaddress,showpacs,byrevtex,floatfix,nofootinbib]{revtex4}
\usepackage{graphicx} 
\usepackage{dcolumn}  
\usepackage{bm}
\usepackage{amsmath}
\usepackage{color}
\usepackage{multirow}

\graphicspath{{ps}}

\renewcommand{\arraystretch}{1.1}

\newcommand{\epem}{\ensuremath{e^{+} e^{-}}}
\newcommand{\qqbar}{\ensuremath{q \bar{q}}}
\newcommand{\BBbar}{\ensuremath{B \bar{B}}}
\newcommand{\BzBzb}{\ensuremath{B^{0} \bar{B}^{0}}}
\newcommand{\BpBm}{\ensuremath{B^{+} B^{-}}}

\newcommand{\pip}{\ensuremath{\pi^{+}}}
\newcommand{\pim}{\ensuremath{\pi^{-}}}

\newcommand{\piz}{\ensuremath{\pi^{0}}}

\newcommand{\Kp}{\ensuremath{K^{+}}}
\newcommand{\Km}{\ensuremath{K^{-}}}

\newcommand{\Ks}{\ensuremath{K^{0}_{S}}}
\newcommand{\Kz}{\ensuremath{K^{0}}}
\newcommand{\Bp}{\ensuremath{B^{+}}}
\newcommand{\Bz}{\ensuremath{B^{0}}}
\newcommand{\Bzb}{\ensuremath{\bar{B}^{0}}}
\newcommand{\Ups}{\ensuremath{\Upsilon(4S)}}

\newcommand{\Brec}{\ensuremath{B^{0}_{\rm Rec}}}
\newcommand{\Btag}{\ensuremath{B^{0}_{\rm Tag}}}

\newcommand{\Mbc}{\ensuremath{M_{\rm bc}}}
\newcommand{\De}{\ensuremath{\Delta E}}

\newcommand{\Lpm}{\ensuremath{{\cal L}^{\pm}_{K/\pi}}}
\newcommand{\Dt}{\ensuremath{\Delta t}}
\newcommand{\Dz}{\ensuremath{\Delta z}}
\newcommand{\FD}{\ensuremath{{\cal F}_{B\bar B/q\bar q}}}
\newcommand{\Hel}{\ensuremath{{\cal H}_{3\pi}}}
\newcommand{\mw}{\ensuremath{M_{3\pi}}}

\newcommand{\taub}{\ensuremath{\tau_{B^{0}}}}
\newcommand{\Dw}{\ensuremath{\Delta w}}
\newcommand{\Dmd}{\ensuremath{\Delta m_{d}}}
\newcommand{\Acpc}{\ensuremath{{\cal A}_{\omega K^{+}}}}
\newcommand{\Acpm}{\ensuremath{{\cal A}_{\omega K^{0}_{S}}}}
\newcommand{\Scpc}{\ensuremath{{\cal S}_{\omega K^{+}}}}
\newcommand{\Scpm}{\ensuremath{{\cal S}_{\omega K^{0}_{S}}}}
\newcommand{\phione}{\ensuremath{\phi_{1}}}

\newcommand{\phithree}{\ensuremath{\phi_{3}}}
\newcommand{\BoK}{\ensuremath{B^{0}\rightarrow \omega K^{0}}}
\newcommand{\BK}{\ensuremath{B \rightarrow \omega K}}
\newcommand{\BoKs}{\ensuremath{B^{0}\rightarrow \omega K^{0}_{S}}}
\newcommand{\BpKp}{\ensuremath{B^{+}\rightarrow \omega K^{+}}}

\begin{document}
\vspace*{-3\baselineskip}

\begin{minipage}[]{0.6\columnwidth}
  \includegraphics[height=3.0cm,width=!]{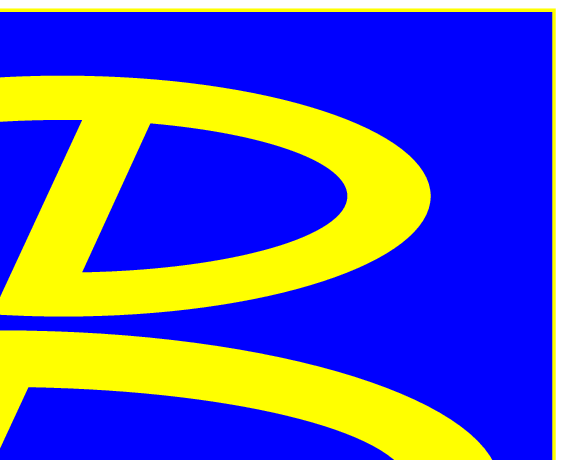}
\end{minipage}
\begin{minipage}[]{0.4\columnwidth}
  \hbox{Belle Preprint 2013-28}
  \hbox{KEK Preprint 2013-56}
\end{minipage}

\title{ \quad\\[1.0cm] Measurement of branching fractions and $\bm{CP}$ violation parameters in $\bm{\BK}$ decays with first evidence of $\bm{CP}$ violation in $\bm{\BoKs}$}

\noaffiliation
\affiliation{University of the Basque Country UPV/EHU, 48080 Bilbao}
\affiliation{Budker Institute of Nuclear Physics SB RAS and Novosibirsk State University, Novosibirsk 630090}
\affiliation{Faculty of Mathematics and Physics, Charles University, 121 16 Prague}
\affiliation{Chiba University, Chiba 263-8522}
\affiliation{University of Cincinnati, Cincinnati, Ohio 45221}
\affiliation{Deutsches Elektronen--Synchrotron, 22607 Hamburg}
\affiliation{Justus-Liebig-Universit\"at Gie\ss{}en, 35392 Gie\ss{}en}
\affiliation{Hanyang University, Seoul 133-791}
\affiliation{University of Hawaii, Honolulu, Hawaii 96822}
\affiliation{High Energy Accelerator Research Organization (KEK), Tsukuba 305-0801}
\affiliation{Indian Institute of Technology Guwahati, Assam 781039}
\affiliation{Indian Institute of Technology Madras, Chennai 600036}
\affiliation{Institute of High Energy Physics, Chinese Academy of Sciences, Beijing 100049}
\affiliation{Institute for High Energy Physics, Protvino 142281}
\affiliation{INFN - Sezione di Torino, 10125 Torino}
\affiliation{Institute for Theoretical and Experimental Physics, Moscow 117218}
\affiliation{J. Stefan Institute, 1000 Ljubljana}
\affiliation{Kanagawa University, Yokohama 221-8686}
\affiliation{Institut f\"ur Experimentelle Kernphysik, Karlsruher Institut f\"ur Technologie, 76131 Karlsruhe}
\affiliation{Korea Institute of Science and Technology Information, Daejeon 305-806}
\affiliation{Korea University, Seoul 136-713}
\affiliation{Kyungpook National University, Daegu 702-701}
\affiliation{\'Ecole Polytechnique F\'ed\'erale de Lausanne (EPFL), Lausanne 1015}
\affiliation{Faculty of Mathematics and Physics, University of Ljubljana, 1000 Ljubljana}
\affiliation{Luther College, Decorah, Iowa 52101}
\affiliation{University of Maribor, 2000 Maribor}
\affiliation{Max-Planck-Institut f\"ur Physik, 80805 M\"unchen}
\affiliation{School of Physics, University of Melbourne, Victoria 3010}
\affiliation{Moscow Physical Engineering Institute, Moscow 115409}
\affiliation{Moscow Institute of Physics and Technology, Moscow Region 141700}
\affiliation{Graduate School of Science, Nagoya University, Nagoya 464-8602}
\affiliation{Kobayashi-Maskawa Institute, Nagoya University, Nagoya 464-8602}
\affiliation{Nara Women's University, Nara 630-8506}
\affiliation{National Central University, Chung-li 32054}
\affiliation{National United University, Miao Li 36003}
\affiliation{Department of Physics, National Taiwan University, Taipei 10617}
\affiliation{H. Niewodniczanski Institute of Nuclear Physics, Krakow 31-342}
\affiliation{Nippon Dental University, Niigata 951-8580}
\affiliation{Niigata University, Niigata 950-2181}
\affiliation{University of Nova Gorica, 5000 Nova Gorica}
\affiliation{Osaka City University, Osaka 558-8585}
\affiliation{Pacific Northwest National Laboratory, Richland, Washington 99352}
\affiliation{Panjab University, Chandigarh 160014}
\affiliation{Peking University, Beijing 100871}
\affiliation{University of Pittsburgh, Pittsburgh, Pennsylvania 15260}
\affiliation{Research Center for Electron Photon Science, Tohoku University, Sendai 980-8578}
\affiliation{University of Science and Technology of China, Hefei 230026}
\affiliation{Seoul National University, Seoul 151-742}
\affiliation{Soongsil University, Seoul 156-743}
\affiliation{Sungkyunkwan University, Suwon 440-746}
\affiliation{School of Physics, University of Sydney, NSW 2006}
\affiliation{Tata Institute of Fundamental Research, Mumbai 400005}
\affiliation{Excellence Cluster Universe, Technische Universit\"at M\"unchen, 85748 Garching}
\affiliation{Toho University, Funabashi 274-8510}
\affiliation{Tohoku Gakuin University, Tagajo 985-8537}
\affiliation{Tohoku University, Sendai 980-8578}
\affiliation{Department of Physics, University of Tokyo, Tokyo 113-0033}
\affiliation{Tokyo Institute of Technology, Tokyo 152-8550}
\affiliation{Tokyo Metropolitan University, Tokyo 192-0397}
\affiliation{Tokyo University of Agriculture and Technology, Tokyo 184-8588}
\affiliation{University of Torino, 10124 Torino}
\affiliation{CNP, Virginia Polytechnic Institute and State University, Blacksburg, Virginia 24061}
\affiliation{Wayne State University, Detroit, Michigan 48202}
\affiliation{Yonsei University, Seoul 120-749}
\affiliation{Max-Planck-Institut f\"ur Physik, 80805 M\"unchen}
\affiliation{Excellence Cluster Universe, Technische Universit\"at M\"unchen, 85748 Garching}
  \author{V.~Chobanova}\affiliation{Max-Planck-Institut f\"ur Physik, 80805 M\"unchen} 
  \author{J.~Dalseno}\affiliation{Max-Planck-Institut f\"ur Physik, 80805 M\"unchen}\affiliation{Excellence Cluster Universe, Technische Universit\"at M\"unchen, 85748 Garching} 
  \author{C.~Kiesling}\affiliation{Max-Planck-Institut f\"ur Physik, 80805 M\"unchen} 
  \author{I.~Adachi}\affiliation{High Energy Accelerator Research Organization (KEK), Tsukuba 305-0801} 
  \author{H.~Aihara}\affiliation{Department of Physics, University of Tokyo, Tokyo 113-0033} 
  \author{D.~M.~Asner}\affiliation{Pacific Northwest National Laboratory, Richland, Washington 99352} 
  \author{V.~Aulchenko}\affiliation{Budker Institute of Nuclear Physics SB RAS and Novosibirsk State University, Novosibirsk 630090} 
  \author{T.~Aushev}\affiliation{Institute for Theoretical and Experimental Physics, Moscow 117218} 
  \author{T.~Aziz}\affiliation{Tata Institute of Fundamental Research, Mumbai 400005} 
  \author{A.~M.~Bakich}\affiliation{School of Physics, University of Sydney, NSW 2006} 
% ...
\collaboration{The Belle Collaboration}
\noaffiliation

\begin{abstract}
  We present a measurement of the branching fractions and charge-parity-($CP$-) violating parameters in \BK\ decays. The results are obtained from the final data sample containing 
  $772 \times 10^{6}$ \BBbar\ pairs collected at the \Ups\ resonance with the Belle detector at the KEKB asymmetric-energy \epem\ collider. We obtain the branching fractions
  $$
  \begin{array}{rcl}
    {\cal B}(\BoK) \!&=&\! (4.5 \pm 0.4 \textrm{ (stat)} \pm 0.3 \textrm{ (syst)})\times 10^{-6},\\
    {\cal B}(\BpKp) \!&=&\! (6.8 \pm 0.4 \textrm{ (stat)} \pm 0.4 \textrm{ (syst)})\times 10^{-6},\\
  \end{array}
  $$
  which are in agreement with their respective current world averages. For the $CP$ violating parameters, we obtain
  $$
  \begin{array}{rcl}
    \Acpm \!&=&\! -0.36 \pm 0.19 \textrm{ (stat)} \pm 0.05 \textrm{ (syst)},\\
    \Scpm \!&=&\! +0.91 \pm 0.32 \textrm{ (stat)} \pm 0.05 \textrm{ (syst)},\\
    \Acpc \!&=&\! -0.03 \pm 0.04 \textrm{ (stat)} \pm 0.01  \textrm{ (syst)},\\
  \end{array}
  $$
  where ${\cal A}$ and $\cal S$ represent the direct and mixing-induced $CP$ asymmetry, respectively. We find no evidence of $CP$ violation in the decay channel \BpKp; however, we obtain the first 
  evidence of $CP$ violation in the \BoKs\ decay channel at the level of 3.1 standard deviations.

\end{abstract}


\maketitle

\tighten

{\renewcommand{\thefootnote}{\fnsymbol{footnote}}}
\setcounter{footnote}{0}

\section{Introduction}
\label{Introduction}
Violation of the combined charge-parity symmetry ($CP$ violation) in the Standard Model (SM) arises from a single irreducible phase in the Cabibbo--Kobayashi--Maskawa (CKM) quark-mixing 
matrix~\cite{Cabibbo,KM}. A primary objective of the Belle experiment is to overconstrain the unitarity triangle of the CKM matrix related to $B_{u,d}$ decays. This permits a 
precision test of the CKM mechanism for $CP$ violation as well as the search for new physics effects. Mixing-induced $CP$ violation in the $B$ sector has been clearly established by 
Belle~\cite{jpsiks_Belle1,jpsiks_Belle2} and BaBar~\cite{jpsiks_BABAR1,jpsiks_BABAR2} in the $\bar b \rightarrow \bar c c \bar s$-induced decay $\Bz \rightarrow J/\psi \Kz$. 

\begin{figure}
  \centering
  \includegraphics[height=120pt,width=!]{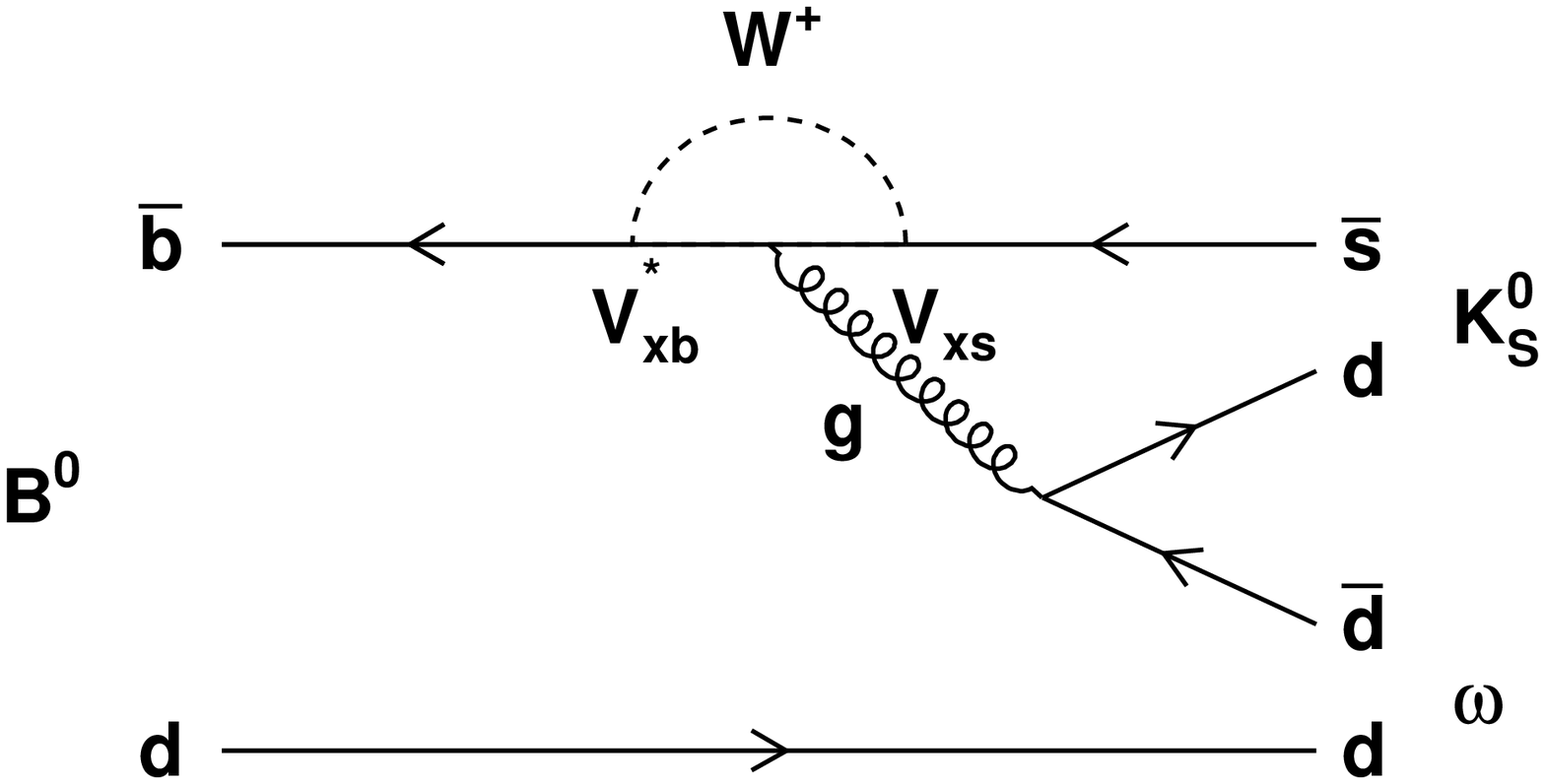}\hspace{0.7cm}
  \includegraphics[height=120pt,width=!]{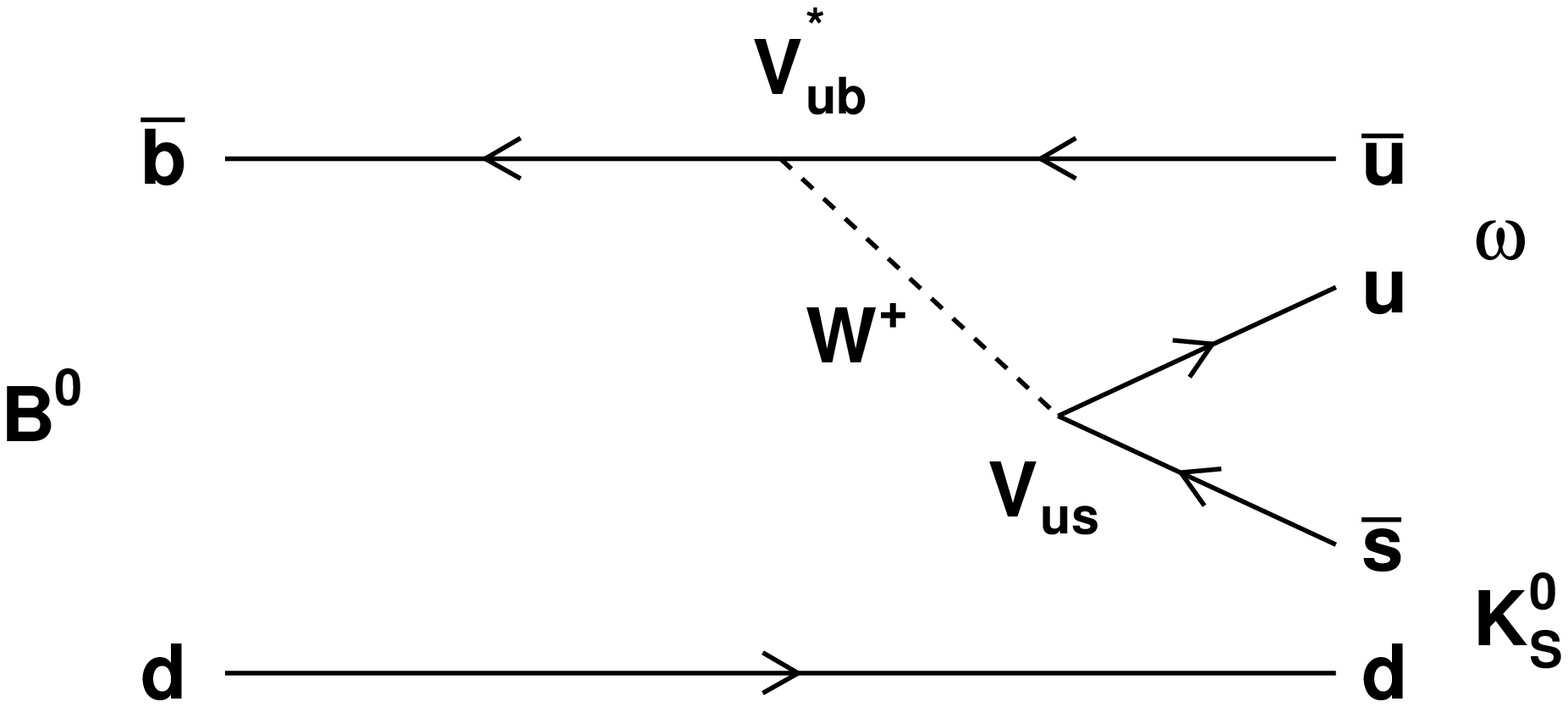}
  \put(-465,95){(a)}
  \put(-218,95){(b)} \\ \vspace{0.4cm}
  \includegraphics[height=120pt,width=!]{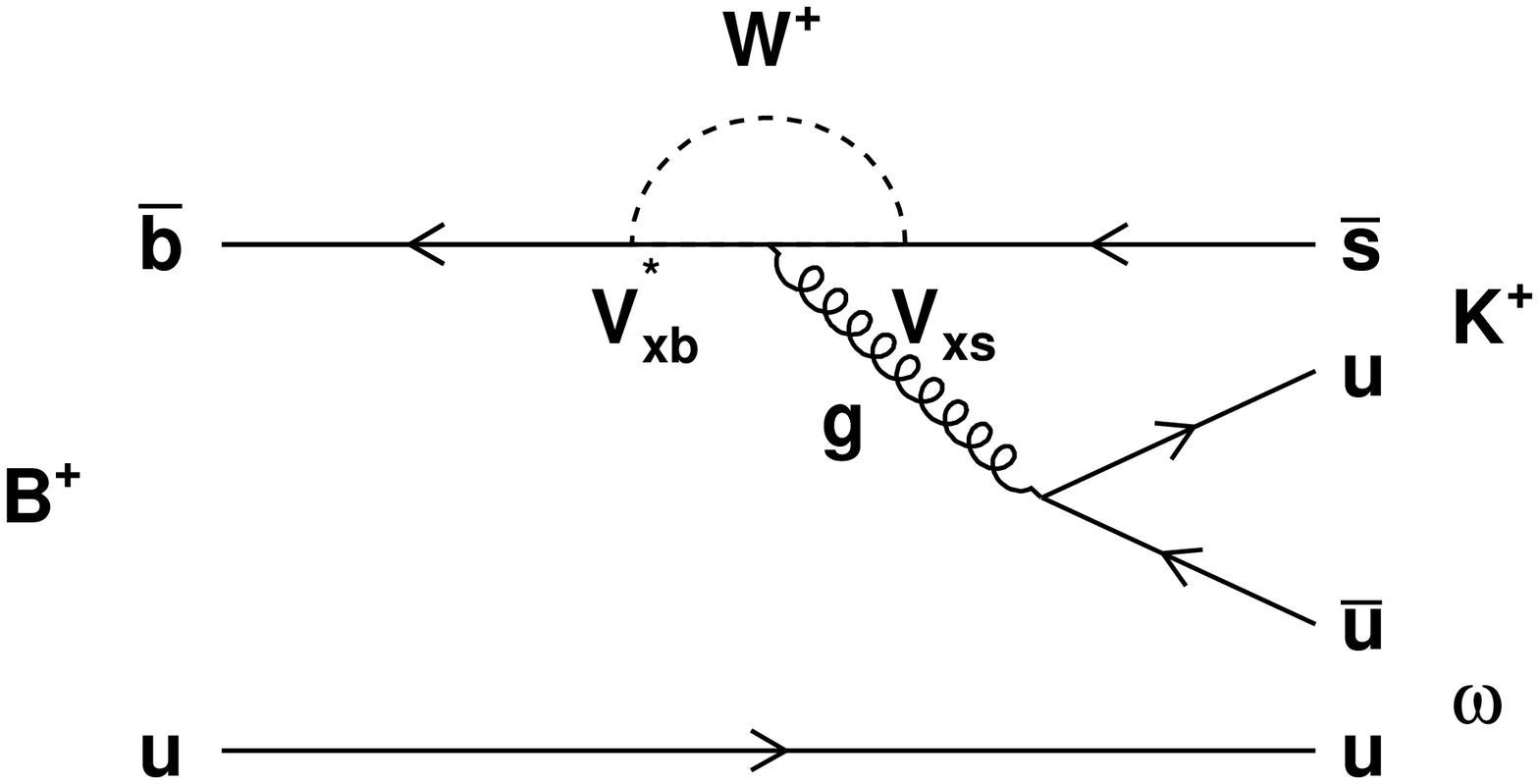}\hspace{0.7cm}
  \includegraphics[height=120pt,width=!]{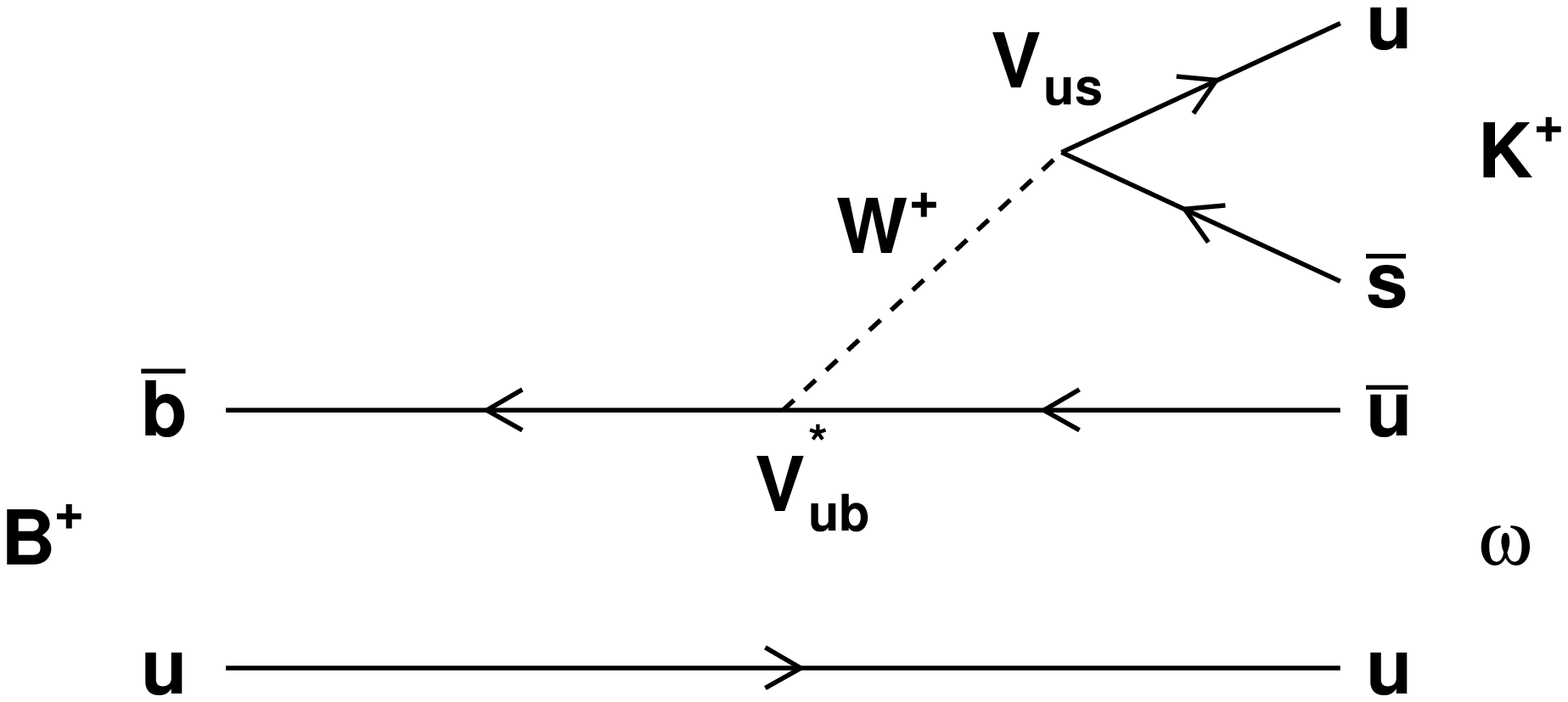}
  \put(-465,95){(c)}
  \put(-218,95){(d)}
  \caption{Leading-order Feynman diagrams for $B \rightarrow \omega K$ decays. For \BoKs, (a) shows the loop (penguin), while (b) shows the tree diagram. 
  For \BpKp, (c) and (d) are the corresponding diagrams. In the penguin diagrams, the subscript $x$ in $V_{xb}$ refers to the flavor of the intermediate-state up-type quark $(x=u,c,t)$.}
  \label{fig_wks}
\end{figure}

Interest has turned toward $b \rightarrow q \bar q s$-mediated decays, where $q$ is a $u$, $d$ or $s$ quark, such as $B\rightarrow \omega(782) K$, for which the physical properties are the subject of 
this paper. These decays proceed predominantly by loop diagrams and are thereby possibly affected by new particles in various extensions of the SM~\cite{NP}. The Feynman diagrams of the neutral and the 
charged decay modes \BoKs\ and \BpKp\ (with charge-conjugate modes included everywhere unless otherwise specified) are shown in 
Fig.~\ref{fig_wks}. The \BoKs\ decays are sensitive to the interior angle of the unitarity triangle $\phione \equiv \arg(-V_{cd}V^{*}_{cb})/(V_{td}V^{*}_{tb})$.
Belle and BaBar have reported measurements on this $CP$-violating phase in this channel~\cite{wks_Belle,wks_BaBar} and other related modes including 
$\Bz \rightarrow \eta'K^{0}$~\cite{etapk0_Belle,wks_BaBar}, $\phi K^{0}_{S}$~\cite{phiks_Belle,phiks_BaBar}, and 
$f_{0}(980)K^{0}_{S}$~\cite{phiks_Belle,phiks_BaBar,f0ks_Belle,f0ks_BaBar1,f0ks_BaBar2}.

The CKM phase \phione\ is accessible experimentally through interference between the direct decay of a $B$ meson to one of the above-mentioned $CP$ eigenstates and \BzBzb\ mixing followed by a decay 
into the same final state. This interference is observable through the time evolution of the decay. We reconstruct \BoKs\ from $\Ups\rightarrow\BzBzb$ decays. As the two
$B$ mesons are produced in  a coherent state, the \Brec\ flavor can be obtained from the other $B$ of the events (\Btag).
The proper time interval between \Brec\ and \Btag, which decay at times $t_{\rm Rec}$ and $t_{\rm Tag}$, 
respectively, is defined as $\Dt \equiv t_{\rm Rec} - t_{\rm Tag}$ measured in the \Ups\ frame. For the coherent \BzBzb\ pairs, the time-dependent decay rate for a $CP$ eigenstate 
when \Btag\ possesses flavor $q$, where \Bz\ has $q=+1$ and \Bzb\ has $q=-1$, is given by
\begin{equation}
  {\cal P}(\Dt, q) = \frac{e^{-|\Dt|/\taub}}{4\taub} \biggl\{1 + q \biggl[\Acpm \cos \Dmd \Dt + \Scpm\sin \Dmd \Dt\biggr]\biggr\}.
\end{equation}
Here, \taub\ is the \Bz\ lifetime, and \Dmd\ is the mass difference between the two mass eigenstates of the neutral $B$ meson. This time dependence assumes $CPT$ invariance 
and that the difference in the decay rates between the two mass eigenstates is negligible. The parameter \Acpm\ measures the direct $CP$ violation, while \Scpm\ is a measure of the amount of 
mixing-induced $CP$ violation. In the limit of a single penguin amplitude with the dominant $t$ quark in the loop, we would expect $\Acpm=0$, $\Acpc=0$, and $\Scpm = \sin{2\phione}$.
However, additional CKM-suppressed contributions carrying different weak phases may not be negligible. As a consequence, direct $CP$ violation can arise, and the measured \Scpm\ 
may differ from $\sin{2\phione}$.

For the charged mode, the direct $CP$ violating parameter \Acpc\ is extracted from the rates of the $\BpKp$ decay, for which the flavor can be 
determined directly from the charge of the kaon on the signal side,
\begin{equation}
  \label{eq_awk}
  \Acpc = \frac{\Gamma(B^{-}\rightarrow \omega K^{-}) - \Gamma(B^{+}\rightarrow \omega K^{+})}{\Gamma(B^{-}\rightarrow \omega K^{-}) + \Gamma(B^{+}\rightarrow \omega K^{+})}.
\end{equation}

\begin{table}
\centering
  \caption{pQCD~\cite{Theory10,Theory12}, QCDF~\cite{Theory4}, SCET 1~\cite{Theory11} and SCET 2~\cite{Theory11} theoretical predictions for the \BoKs\ and \BpKp\ branching fractions 
  (in units of $10^{-6}$) and $CP$ parameters (in units of $10^{-2}$). The meaning of each uncertainty for these approaches is given in the corresponding references.}
  \begin{tabular}
    {@{\hspace{0.4cm}}l@{\hspace{0.4cm}}  @{\hspace{0.4cm}}r@{\hspace{0.4cm}}  @{\hspace{0.4cm}}r@{\hspace{0.4cm}}  @{\hspace{0.4cm}}r@{\hspace{0.4cm}} @{\hspace{0.4cm}}r@{\hspace{0.4cm}}}
    \hline \hline
    Parameter & pQCD & QCDF & SCET 1 & SCET 2\\
    \hline
    ${\cal B}(\BoK)$  	& $9.8^{+8.6+6.7}_{-4.9-4.3}$ 	& $4.1^{+4.2+3.3}_{-1.7-2.2}$ 		& $4.1^{+2.1+0.8}_{-1.7-0.7}$ 		& $4.9^{+1.9+0.7}_{-1.6-0.6}$\\
    ${\cal B}(\BpKp)$ 	& $10.6^{+10.4+7.2}_{-5.8-4.4}$	& $4.8^{+4.4+3.5}_{-1.9-2.3}$ 		& $5.1^{+2.4+0.9}_{-1.9-0.8}$ 		& $5.9^{+2.1+0.8}_{-1.7-0.7}$\\
    \Acpm 	      	& $-3^{+2+2}_{-4-3}$	& $-4.7^{+1.8+5.5}_{-1.6-5.8}$ 		& $5.2^{+8.0+0.6}_{-9.2-0.6}$		& $3.8^{+5.2+0.3}_{-5.4-0.3}$\\
    \Scpm             	& $84^{+3+0}_{-7-2}$	& $84^{+5+4}_{-5-6}$			& $51^{+5+2}_{-6-2}$			& $80^{+2+1}_{-2-1}$\\
    \Acpc             	& $32^{+15+4}_{-17-5}$	& $22.1^{+13.7+14.0}_{-12.8-13.0}$	& $11.6^{+18.2+1.1}_{-20.4-1.1}$ 	& $12.3^{+16.6+0.8}_{-17.3-1.1}$\\
    \hline \hline
  \end{tabular}
  \label{tab_th_pred}
\end{table}

The measurements of the branching fractions and $CP$ parameters in \BK\ decays provide an important test of the QCD factorization (QCDF), perturbative QCD (pQCD) and soft collinear 
effective theory (SCET) approaches. The predictions made by these SM-based theoretical calculations are summarized in Table~\ref{tab_th_pred}.
These approaches predict a relatively sizeable direct $CP$ asymmetry in \BpKp\ and expect \Scpm\ to be slightly higher than in $b \rightarrow c \bar c s$ decays. 
However, current experimental measurements of \Scpm~\cite{wks_Belle,wks_BaBar} could indicate the opposite, motivating more precise experimental determinations to reduce the large statistical 
uncertainties. Finally, combinations of measurements of the branching fractions and charge asymmetries of charmless $B$ meson decays can be used in phenomenological fits to understand the 
relative importance of tree and penguin contributions and may provide sensitivity to the CKM angle $\phithree \equiv \arg(-V_{ud}V^{*}_{ub})/(V_{cd}V^{*}_{cb})$~\cite{Phen2,Phen4}.



Experimentally, clear signals have been observed by Belle and BaBar for 
\BoK\ and \BpKp\ with similar branching fractions~\cite{wk_Belle,wk_BaBar}. 
Measurements of the $CP$ violation parameters and the branching fractions for these channels were reported by Belle and BaBar. All previous measurements are summarized in Table~\ref{tab_wks_prev}.

In this paper, we present an updated measurement of the branching fractions and $CP$ violation parameters in \BK\ decays using the full Belle data set with $772 \times 10^{6}$ \BBbar\ pairs; 
this supersedes the previous Belle analysis. In Sec.~\ref{Data Set And Belle Detector}, we briefly describe the data set and Belle detector. 
The selection criteria used to identify signal candidates and suppress backgrounds along with the definition of the variables that will be used to extract the physical signal 
parameters are explained in Sec.~\ref{Event Selection}. Following that, the signal and background models for these variables will be discussed in Sec.~\ref{Event Model}. In Sec.~\ref{Results}, the results 
of the fit are presented along with a discussion of the systematic uncertainties in Sec.~\ref{Systematic Uncertainties}. Finally, our conclusions are given in Sec.~\ref{Conclusion}.

\begin{table}
  \centering
  \caption{Summary of $B\rightarrow \omega K$ branching fractions (in units of $10^{-6}$) and $CP$ violation parameters (in units of $10^{-2}$) obtained by Belle~\cite{wk_Belle,wks_Belle} and BaBar~\cite{wk_BaBar,wks_BaBar}. For all parameters, the first uncertainty is statistical and the second is systematic.}
  \begin{tabular}
    {@{\hspace{0.4cm}}c@{\hspace{0.25cm}}  @{\hspace{0.25cm}}c@{\hspace{0.4cm}}  @{\hspace{0.25cm}}c@{\hspace{0.4cm}}  @{\hspace{0.4cm}}c@{\hspace{0.25cm}}
    @{\hspace{0.25cm}}c@{\hspace{0.4cm}}}
    \hline \hline
    Parameter & Belle & Belle & BaBar \\
    & ($388\times10^6$ \BBbar\ pairs) & ($535\times10^6$ \BBbar\ pairs) & ($467\times10^6$ \BBbar\ pairs) \\
    \hline
    ${\cal B}(\BoKs)$ & $4.4^{+0.8}_{-0.7}\pm0.4$ & - & $5.4\pm0.8\pm0.3$ \\
    $\Acpm$ & - & $-9\pm29\pm6$ & $52^{+20}_{-22}\pm3$ \\
    $\Scpm$ & - & $11\pm46\pm7$ & $55^{+26}_{-29}\pm2$ \\
    \hline \hline
  \end{tabular}
  \vspace{0.5cm}
  
  \begin{tabular}
    {@{\hspace{0.5cm}}c@{\hspace{0.25cm}}  @{\hspace{0.25cm}}c@{\hspace{0.5cm}}  @{\hspace{0.25cm}}c@{\hspace{0.5cm}}  @{\hspace{0.25cm}}c@{\hspace{0.5cm}}}
    \hline \hline
    Parameter & Belle & BaBar\\
    & ($388\times10^6$ \BBbar\ pairs) & ($383\times10^6$ \BBbar\ pairs)\\
    \hline
    ${\cal B}(\BpKp)$ & $8.1\pm0.6\pm0.6$ & $6.3\pm0.5\pm0.3$ \\
    $\Acpc$ & $5^{+8}_{-7}\pm1$ & $-1\pm7\pm1$ \\
    \hline \hline
  \end{tabular}  
  \label{tab_wks_prev}
\end{table}

\section{Data set and Belle detector}
\label{Data Set And Belle Detector}
The results in this paper are based on the \Ups\ final data sample containing $772 \times 10^{6}$ \BBbar\ pairs collected with the Belle detector at the KEKB asymmetric-energy \epem\ 
($3.5$ on $8~{\rm GeV}$) collider~\cite{KEKB}. At the \Ups\ resonance ($\sqrt{s}=10.58$~GeV), the Lorentz boost of the produced \BBbar\ pairs is $\beta\gamma =0.425$ 
nearly along the $+z$ direction, which is opposite the positron beam direction.

The Belle detector is a large-solid-angle magnetic spectrometer that consists of a silicon vertex detector (SVD), a 50-layer central drift chamber (CDC), an array of
aerogel threshold Cherenkov counters (ACC),  
a barrellike arrangement of time-of-flight scintillation counters (TOF), and an electromagnetic calorimeter (ECL) comprising CsI(Tl) crystals located inside 
a superconducting solenoid coil that provides a 1.5~T magnetic field.  An iron flux return located outside of the coil is instrumented to detect $K_L^0$ mesons and to identify
muons.  The detector is described in detail elsewhere~\cite{Belle}.
Two inner detector configurations were used. A 2.0-cm-radius beampipe and a three-layer silicon vertex detector (SVD1) were used for the first sample
of $152 \times 10^6 B\bar{B}$ pairs, while a 1.5-cm-radius beampipe, a four-layer silicon detector (SVD2), and a small-cell inner drift chamber were used to record  
the remaining $620 \times 10^6 B\bar{B}$ pairs~\cite{svd2}. We use a GEANT-based Monte Carlo (MC) simulation to model the response of the detector and to determine 
its acceptance~\cite{GEANT}.

\section{Event selection}
\label{Event Selection}

\subsection{$B$ candidate selection}

We reconstruct $B^{0}$ ($B^{+}$) meson candidates from an $\omega$ and a \Ks\ ($K^{+}$) candidate. The $\omega$ candidates are reconstructed from $\pip\pim\piz$ combinations, where 
$\piz$ is reconstructed from two photons.
Charged tracks forming an $\omega$ candidate and the prompt kaon track are identified using a loose requirement on the distance of closest approach with respect to 
the interaction point (IP) along the beam direction, $|dz| < 4.0 \; {\rm cm}$, and in the transverse plane, $dr < 0.4 \; {\rm cm}$. Additional SVD requirements of at least two $z$ hits and one 
$r-\phi$ hit~\cite{ResFunc} are imposed on at least one charged track forming an $\omega$ candidate so that a good quality vertex of the reconstructed $B$ meson candidate can be determined. 
Using information obtained from the CDC, ACC, and TOF, particle identification (PID) is determined from a likelihood ratio ${\cal L}_{i/j} \equiv {\cal L}_{i}/({\cal L}_{i} + {\cal L}_{j})$. 
Here, ${\cal L}_{i}$ (${\cal L}_{j}$) is the likelihood that the particle is of type $i$ ($j$). To suppress background due to electron misidentification, ECL information is used to veto 
particles consistent with the electron hypothesis~\cite{eid}. The PID ratios of all charged tracks \Lpm\ are used to identify them as a kaon or a pion. 

A \Ks\ candidate consists of two oppositely charged $\pi$ candidates. We only consider \Ks\ candidates with vertices displaced from the IP; the displacement depends on the \Ks\ momentum. 
Only events in the mass window $|M_{\pi\pi} - m_{\Ks}| < 16 \; {\rm MeV}/c^{2}$ are accepted, where $m_{\Ks}$ is the world average mass of the \Ks~\cite{PDG}. 

Photons ($\gamma$) are identified as isolated 
ECL clusters that are not matched to any charged particle track. To suppress the combinatorial background, the photons are required to have a minimum energy of $E_{\gamma} > 50\; {\rm MeV}$ in 
the ECL barrel region and  $E_{\gamma} > 100\; {\rm MeV}$ in the ECL end cap regions, where the barrel region covers the polar angle range $32^{\circ} < \theta < 130^{\circ}$ and the end cap 
regions cover the forward and backward regions in the ranges $12^{\circ} < \theta < 32^{\circ}$ and $130^{\circ} < \theta < 157^{\circ}$.
Two $\gamma$ candidates are combined to form a \piz\ candidate that must satisfy $|M_{\gamma \gamma} - m_{\piz}| < 16\; {\rm MeV}/c^{2}$, where $m_{\piz}$ is the world average
mass of the \piz~\cite{PDG}.  

An $\omega$ candidate consists of two oppositely charged pion candidates and a \piz\ candidate with the 
requirement $|\mw - m_{\omega}| < 50 \; {\rm MeV}/c^{2}$, where $m_{\omega}$ is the world average mass of the $\omega$~\cite{PDG}. The mass distribution of the $\omega$ candidates is shown in 
Figs.~\ref{fig_data_wks}d and~\ref{fig_data_wkp}d. The mass cuts given above correspond to about three times the typical experimental resolution of the \Ks, \piz, and $\omega$ world average masses. We also reconstruct the cosine of the helicity angle \Hel\ of the $\omega$ candidates. This angle is defined as that between 
the direction of \Brec\ and the normal to the three-pion decay plane, both calculated in the rest frame of the $\omega$ candidate. The \Hel\ distributions for all components are presented in 
Figs.~\ref{fig_data_wks}e and~\ref{fig_data_wkp}e.

Reconstructed $B$ meson candidates are identified with two nearly uncorrelated kinematic variables: the beam-energy-constrained mass 
$\Mbc \equiv \sqrt{(E^{\rm CMS}_{\rm beam})^{2} - (p^{\rm CMS}_{B})^{2}}$ and the energy difference $\De \equiv E^{\rm CMS}_{B} - E^{\rm CMS}_{\rm beam}$, where $E^{\rm CMS}_{\rm beam}$ 
is the beam energy and $E^{\rm CMS}_{B}$ ($p^{\rm CMS}_{B}$) is the energy (momentum) of the $B$ meson, all evaluated in the $e^+ e^-$ center-of-mass system (CMS). The $B$ meson candidates 
that satisfy $\Mbc > 5.25 \; {\rm GeV}/c^{2}$ and $-0.15 \; {\rm GeV} < \De < 0.1 \; {\rm GeV}$ are retained for further analysis. The distributions of these two variables are shown in 
Figs.~\ref{fig_data_wks}a and~\ref{fig_data_wkp}a and Figs.~\ref{fig_data_wks}b and~\ref{fig_data_wkp}b, respectively.

\subsection{$B$ vertex reconstruction}

As the \Brec\ and \Btag\ are almost at rest in the CMS, the difference in decay time between the two $B$ meson candidates, $\Delta t$, can be determined approximately from the displacement 
in $z$ between the final-state decay vertices as
\begin{equation}
  \Dt \simeq \frac{(z_{\rm Rec} - z_{\rm Tag})}{\beta \gamma c} \equiv \frac{\Delta z}{\beta \gamma c},
\end{equation}
where $\beta\gamma=0.425$ is the Lorentz boost of the \Ups\ in the lab frame and $c$ is the speed of light.

The \Brec\ decay vertex
is determined from one or two charged daughters of the $\omega$, depending on whether they pass the SVD requirements.
A single-track vertex is possible as an IP constraint using the known beam profile in the $x$-$y$ plane is always included as a pseudotrack in the vertex finding algorithm.
To obtain the \Dt\ distribution, we also reconstruct the tag-side vertex from the tracks not used 
to reconstruct \Brec~\cite{ResFunc}. Candidate events must satisfy the loose requirements $|\Dt| < 70 \; {\rm ps}$ and $h_{\rm Rec, Tag} < 50$, where $h_{\rm Rec, Tag}$ is the multitrack vertex 
goodness of fit, calculated in three-dimensional space without the IP profile constraint~\cite{jpsiks_Belle2}. To avoid the necessity of also modeling the event-dependent observables that 
describe the \Dt\ resolution in the fit~\cite{punzi}, the vertex uncertainty is required to satisfy $\sigma^{\rm Rec, Tag}_z < 200 \; \mu {\rm m}$ for multitrack vertices and 
$\sigma^{\rm Rec, Tag}_z < 500 \; \mu {\rm m}$ for vertices reconstructed from single tracks and the IP constraint.
The efficiency of the vertexing algorithm is 91\%.

\subsection{Flavor tagging}

The \Btag\ flavor is determined from the remaining tracks and photons left over from the \Brec\ reconstruction. The flavor-tagging procedure is described in Ref.~\cite{Tagging}. The tagging information is represented by two parameters, 
the \Btag\ flavor $q$ and the flavor-tagging quality $r$. The parameter $r$ is continuous and determined on an event-by-event basis with an algorithm trained on MC-simulated events, ranging from
zero for no flavor discrimination to unity for an unambiguous flavor 
assignment. To obtain a data-driven replacement for $r$, we divide its range into seven regions and determine a probability of mistagging $w$ for each $r$ region using high statistics control samples.
If MC describes the data perfectly, then $r=1-2w$. The $CP$ asymmetry in data is thus diluted by a factor $1-2w$ instead of the MC-determined $r$.
The measure of the flavor-tagging algorithm performance is the total effective tagging efficiency $\epsilon_{\rm eff} = (1-2w)^2\epsilon_{\rm Tag}$, rather than the raw tagging efficiency 
$\epsilon_{\textrm{Tag}}$, as the statistical significance of the $CP$ parameters is proportional to $(1-2w)\sqrt{\epsilon_{\rm Tag}}$. These are determined from data to be 
$\epsilon_{\rm eff} = 0.284\pm0.010$ and $\epsilon_{\rm eff} = 0.301\pm0.004$ for the SVD1 and SVD2 data, respectively~\cite{jpsiks_Belle2}. After the flavor-tagging algorithm has 
been applied, 99.8\% of all signal candidates remain.

\subsection{Continuum reduction}

The dominant background in the reconstruction of \Brec\ arises from \qqbar\ (continuum) events, where $q = u,d,s,c$. Since their topology tends to be jetlike in contrast to the spherical 
\BBbar\ decay, continuum events are suppressed with a Fisher discriminant~\cite{nazi_stuff} based on modified Fox--Wolfram moments~\cite{KSFW}. The \BBbar\ training sample is taken from signal 
MC simulation, while the \qqbar\ training sample is based on sideband data taken at the \Ups\ resonance with minimal contamination from $B$ mesons, 
defined as $\Mbc < 5.25 \; {\rm GeV}/c^{2}$ and $0.05 \leq \De \leq 0.2 \; {\rm GeV}$. The distributions of both samples are combined to form a likelihood ratio
distribution, which peaks at unity for \BBbar\ events and at zero for \qqbar\ background. 
To further improve the signal-background distinction, the likelihood ratio of the Fisher discriminant is multiplied by the likelihood ratio of
the polar angle of the $B$ meson candidate in the CMS, $\cos \theta_{B}$, which follows a $1-\cos^{2} \theta_{B}$ distribution for \BBbar\ events while being flat for the 
continuum. We employ a loose selection on the resulting likelihood ratio, ${\cal{L}}_{\BBbar/\qqbar}\geq0.2$, which reduces \qqbar\ background by 
62\% with a signal efficiency of 94\%. To make the likelihood ratio distribution easier to parametrize, it is transformed into a Gaussian-like distribution according to 
\begin{equation}
\FD = \log{\frac{{\cal{L}}_{\BBbar/\qqbar}-0.2}{1-{\cal{L}}_{\BBbar/\qqbar}}}.
\label{e_Lsb_trans}
\end{equation}
The signal and background \FD\ distributions are shown in Figs.~\ref{fig_data_wks}c and~\ref{fig_data_wkp}c. 

\subsection{Reconstruction efficiency}

After these selection criteria, we obtain from signal MC the detection efficiencies $\epsilon^{s,d}$ for each SVD configuration $s$ and decay channel $d$. These are summarized in Table~\ref{tab_eff_pid}. 
The uncertainties come from limited MC simulation statistics. We also determine correction factors to the efficiency $\eta^{s,d}$ that account for the difference between data and MC simulation as 
calculated by independent studies at Belle. These correction factors in 
our reconstruction algorithm arise only from PID and are determined from an inclusive $D^{∗+}\rightarrow D^0[K^{-} \pip]\pip$ sample. They are summarized in Table~\ref{tab_eff_pid}. 

About 15\% of all events have more than one $B$ candidate. For these events, an arbitrary candidate is selected. From MC simulation, 2\% of signal events is found to be 
misreconstructed, defined as being events where at least one of the tracks entering the vertex reconstruction does not belong to the $B$ meson of interest. 

\begin{table}
  \centering
  \caption{Summary of the detection efficiencies (eff.) (top) and PID correction factors (bottom) for \BoKs\ and \BpKp. The values of the efficiency yields and their uncertainties are obtained from signal 
  MC.}
  \begin{tabular}
    {@{\hspace{0.5cm}}l@{\hspace{0.5cm}} @{\hspace{0.5cm}}l@{\hspace{0.5cm}}  @{\hspace{0.5cm}}l@{\hspace{0.5cm}}}
    \hline \hline 
    Decay & Eff. SVD1 ($\epsilon^{1,d}$) & Eff. SVD2 ($\epsilon^{2,d}$)\\
    \hline
    \BoKs & $0.1136 \pm 0.0003$ & $0.1454 \pm 0.0004$\\
    \BpKp & $0.1828 \pm 0.0004$ & $0.2195 \pm 0.0005$\\
    \hline \hline
  \end{tabular}
  
 \vspace{0.5cm}
 
  \begin{tabular}
    {@{\hspace{0.5cm}}l@{\hspace{0.5cm}} @{\hspace{0.5cm}}l@{\hspace{0.5cm}}  @{\hspace{0.5cm}}l@{\hspace{0.5cm}}}
    \hline \hline 
    Decay & PID SVD1 ($\eta^{1,d}$) & PID SVD2 ($\eta^{2,d}$)\\
    \hline
    \BoKs & $0.961 \pm 0.010$ & $0.959 \pm 0.020$\\
    \BpKp & $0.948 \pm 0.018$ & $0.923 \pm 0.028$\\
    \hline \hline
  \end{tabular}
  \label{tab_eff_pid}
\end{table}

\section{Event model}
\label{Event Model}

The two branching fractions and three $CP$ violation parameters of \BK\ are extracted from a sequence of seven-dimensional unbinned extended maximum likelihood fits to 
\De, \Mbc, \FD, \mw, \Hel, \Dt, and $q$ performed simultaneously on the two data samples $d$, each divided into seven bins ($l = 0..6$) in the flavor-tag quality $r$ and two SVD configurations $s$. 
In the first fit, the two branching fractions and $CP$ parameters of the neutral mode are determined. In two further fits, the charged data sample is divided into two 
subsambles depending on the $B$ charge. From these, two signal yields are extracted in order to determine \Acpc\ according to Eq. (~\ref{eq_awk}).
The following categories are considered in the event model: signal, misreconstructed signal, continuum, charm and charmless neutral and charged $B$ meson decays, and charm and charmless peaking 
backgrounds. The probability density function (PDF) for each category is usually taken as the product of PDFs for each variable (unless otherwise stated), 
${\cal P}^{l,s,d}(\De,\Mbc,\FD,\mw,\Hel,\Dt,q) \equiv {\cal P}^{l,s,d}(\De) \times {\cal P}^{l,s,d}(\Mbc) \times {\cal P}^{l,s,d}(\FD) \times {\cal P}^{l,s,d}(\mw) \times {\cal P}^{l,s,d}(\Hel) 
\times {\cal P}^{l,s,d}(\Dt,q)$, in each $l,s,d$ bin as most correlations between the fit observables are negligible. We describe these fit models for each category explicitly in the following 
subsections.

\subsection{Signal model}
\label{Signal Model}

The signal shapes are determined from the signal MC events where the $\pi^{+}$ and $\pi^{-}$ forming the $\omega$ candidate are correctly reconstructed. The PDF for \De\ is the sum of three Gaussian functions and a linear function,
\begin{eqnarray}
  {\cal P}^{l,s,d}_{\rm Sig}(\De)&\equiv& f^{s,d}_{1}G(\De;\, \mu^{s,d}_{1}+\mu^s_{C}, \, \sigma^{s,d}_{1}\sigma^s_{C})  \nonumber \\
  &&+ f^{s,d}_{\rm2}G(\De; \, \mu^{s,d}_{2} +\mu^{s,d}_{1} + \mu^s_{C},\, \sigma^{s,d}_{2}\sigma^{s,d}_{1}\sigma^s_{C}) \nonumber \\
  &&+ f^{s,d}_{\rm3}G(\De;\, \mu^{s,d}_{3} + \mu^{s,d}_{2}+\mu^{s,d}_{1}+\mu^s_{C}, \, \sigma^{s,d}_{3}\sigma^{s,d}_{2}\sigma^{s,d}_{1}\sigma^s_{C}) \nonumber \\
  &&+ (1-f^{s,d}_{1}-f^{s,d}_{2}-f^{s,d}_{3}) (1 + c^{s,d} \De ),
  \label{e_desig}
\end{eqnarray}
where the two-tail Gaussians are parametrized relative to the main Gaussian. This PDF also incorporates calibration parameters $\mu^s_{C}$ and $\sigma^s_{C}$, which correct for the difference between 
data and MC simulation. These parameters calibrate the mean and width of the main Gaussian component and are the only parameters shared between both data samples. They are fixed to zero and unity, 
respectively, in the fit to determine the signal model from MC, but are free in the fit to data. 
Because of the low signal yield of the neutral mode, we determine the calibration factors in a simultaneous fit of two decay channels instead of extracting them from a separate control sample fit. Thus, 
to first order, we do not need to consider the related systematic uncertainties that arise from the difference between data and MC.
Because of our definition of correctly reconstructed events, the linear part of the PDF is necessary to describe events where the \piz\ or $K_{S}^{0}$ are incorrectly reconstructed.

The PDF for \Mbc\ is taken to be the sum of three Gaussians and an empirically determined shape referred to as an ARGUS function ($A$)~\cite{ARGUS}, which has an event-dependent cutoff at 
$E^{\rm CMS}_{\rm beam}$. The ARGUS function represents events analogous to those of the linear function models in \De. Because of correlations of 4-5\% between \De\ and \Mbc\ in the different 
data samples, the dependency of the main mean and relative fraction of the main Gaussian leads to the parametrization
\begin{eqnarray}
  {\cal P}^{l,s,d}_{\rm Sig}(\Mbc|\De)&\equiv& (f^{s,d}_{1}+\alpha^{s,d} |\De|)G(\Mbc;\, \mu^{s,d}_{1}+\mu^s_{C}+\beta^{s,d} \De, \, \sigma^{s,d}_{1}\sigma^s_{C})  \nonumber \\
  &&+ f^{s,d}_{2}G(\Mbc; \, \mu^{s,d}_{2} +\mu^{s,d}_{1} + \mu^s_{C}+\beta^{s,d} \De,\, \sigma^{s,d}_{2}\sigma^{s,d}_{1}\sigma^s_{C}) \nonumber \\
  &&+ f^{s,d}_{3}G(\Mbc;\, \mu^{s,d}_{3} + \mu^{s,d}_{2}+\mu^{s,d}_{1}+\mu^s_{C}+\beta^{s,d} \De, \, \sigma^{s,d}_{3}\sigma^{s,d}_{2}\sigma^{s,d}_{1}\sigma^s_{C}) \nonumber \\
  &&+ (1-[f^{s,d}_{1}+\alpha^{s,d} |\De|]-f^{s,d}_{2}-f^{s,d}_{3}) A(\Mbc; a^{s,d}),
\end{eqnarray}
where $\alpha^{s,d}$ and $\beta^{s,d}$ represent the additional correlation parameters and $a^{s,d}$ is the shape parameter of the ARGUS function. As in \De, only the shared calibration parameters are free in the fit to data.

The PDF for \FD\ is taken to be the sum of three Gaussians in each flavor-tag bin $l$,
\begin{eqnarray}
  {\cal P}^{l,s,d}_{\rm Sig}(\FD)&\equiv& f^{l,s,d}_{1}G(\FD;\, \mu^{l,s,d}_{1}+\mu^{l,s}_{C}, \, \sigma^{l,s,d}_{1}\sigma^{l,s}_{C})  \nonumber \\
  &&+ f^{l,s,d}_{\rm2}G(\FD; \, \mu^{l,s,d}_{2} +\mu^{l,s,d}_{1} + \mu^{l,s}_{C},\, \sigma^{l,s,d}_{2}\sigma^{l,s,d}_{1}\sigma^{l,s}_{C}) \nonumber \\
  &&+ f^{s,d}_{\rm3}G(\FD;\, \mu^{l,s,d}_{3} + \mu^{l,s,d}_{2}+\mu^{l,s,d}_{1}+\mu^{l,s}_{C}, \, \sigma^{l,s,d}_{3}\sigma^{l,s,d}_{2}\sigma^{l,s,d}_{1}\sigma^{l,s}_{C}),
\label{eq:fd}
\end{eqnarray}
This time, the shared calibration parameters also depend on $l$ and are free in the fit to data.

The \mw\ PDF also consists of the sum of three Gaussians where the correlation of 27\% between \De\ and \mw\ is considered as
\begin{eqnarray}
  {\cal P}^{l,s,d}_{\rm Sig}(M_{3\pi}| \,\De) &\equiv& 
  f^{s,d}_{1}G(M_{3\pi};\, \mu^{s,d}_{1} + \mu^s_{C} + \alpha^{s,d}\De, \,\sigma^{s,d}_{1}\sigma^s_{C}+\beta^{s,d}\De^{2}) 
  \nonumber \\
  &&+f^{s,d}_{2}G(M_{3\pi};\, \mu^{s,d}_{2}+\mu^{s,d}_{1} + \mu^s_{C} + \alpha^{s,d}\De ,\, \sigma^{s,d}_{2}[\sigma^{s,d}_{1}\sigma^s_{C}+\beta^{s,d}\De^{2}]) \nonumber \\
  &&  + (1-f^{s,d}_{1}-f^{s,d}_{2}) G(M_{3\pi};\mu^{s,d}_{3} +\mu^{s,d}_{2} + \mu^{s,d}_{1} + \mu^s_{C}+ \alpha^{s,d}\De, \nonumber\\
  && \hspace{133pt}\sigma^{s,d}_{3}\sigma^{s,d}_{2}[\sigma^{s,d}_{1}\sigma^s_{C}+\beta^{s,d}\De^{2}]),
  \label{e_dwmsig}
\end{eqnarray}
where $\alpha^{s,d}$ and $\beta^{s,d}$ are the correlation parameters.

The \Hel\ shape is modeled with the sum of symmetric Chebyshev 
polynomials $C_{i}$, up to fourth order and is determined from MC:
\begin{eqnarray}
  {\cal P}^{l,s,d}_{\rm Sig}(\Hel) \equiv 1+\sum_{i=1}^{2}c^{s,d}_{2i}C_{2i}(\Hel).
\label{eq:hel}
\end{eqnarray}
The PDF of \Dt\ and $q$ for \BoKs\ is given by
\begin{eqnarray}
  {\cal P}^{l,s, \omega \Ks}_{\rm Sig}(\Dt, q) \equiv (1-f^{s}_{\rm Out})\,\frac{e^{-|\Dt|/\taub}}{4\taub} &\biggl\{&1-q\Dw^{l,s}+q(1-2w^{l,s})\times 
   \biggl[\Acpm\cos \Dmd \Dt \nonumber \\
     & &+ \Scpm \sin \Dmd \Dt\biggr]\biggr\} \otimes R^{s}_{\BzBzb}(\Dt) \nonumber\\
   \hspace{-80pt} + f^{s}_{\rm Out}\, \frac{1}{2}G(\Dt; 0, \sigma^{s}_{\rm Out}),
\label{eq:dtq}
\end{eqnarray}
which accounts for $CP$ dilution from the probability of incorrect flavor tagging $w^{l,s}$ and the wrong tag difference $\Dw^{l,s}$ between \Bz\ and \Bzb. The values of $w^{l,s}$ and $\Dw^{l,s}$ 
are determined from flavor-specific control samples using the method as described in Ref~\cite{Tagging}. The physics parameters \taub\ and \Dmd\ are fixed to their respective current world averages~\cite{PDG}. 
This PDF is convolved with the \Dt\ resolution function for neutral $B$ particles $R^{s}_{\BzBzb}$, as given in Ref.~\cite{jpsiks_Belle2}, which describes \Dt\ smearing effects due to detector 
resolution, the use of nonprimary tracks to form the tag-side vertex, and the kinematic approximation of calculating \Dt\ from the one-dimensional separation \Dz. To account for the possibility of 
remaining outlier \Dt\ events that cannot be described by the \Dt\ resolution function, a broad Gaussian centered at zero is introduced with a relative fraction 
$f^{s}_{\rm Out}$ and width $\sigma^{s}_{\rm Out}$ parameters given in Ref.~\cite{jpsiks_Belle2}. For \BpKp, the PDF is given by
\begin{equation}
  {\cal P}^{l,s, \omega \Kp}_{\rm Sig}(\Dt, q) \equiv (1-f^{s}_{\rm Out})\, \frac{e^{-|\Dt|/\tau_{B^+}}}{4\tau_{B^+}} \otimes R^{s}_{\BpBm}(\Dt)  + f^{s}_{\rm Out}\, \frac{1}{2}G(\Dt; 0, \sigma^{s}_{\rm Out}),
\label{eq:dtq_wk}
\end{equation}
where ${\cal R}^{s}_{\BpBm}$ is the \Dt\ resolution function for charged $B$ meson decays~\cite{jpsiks_Belle2}. There are no free
parameters in this case.

\subsection{Misreconstructed signal model}
\label{Misreconstructed Signal Model}
The misreconstructed model shape is determined from signal MC simulation events with an incorrectly reconstructed vertex.
The PDFs for \De\ and \mw\ are the sum of a Gaussian and a linear function, while the \Mbc\ PDF is a combination of an asymmetric Gaussian and an ARGUS function. The shape of \FD\ is the same as 
Eq. (\ref{eq:fd}) from the signal model and shares most of the parameters including calibration factors; however, the main mean in each flavor-tag bin is determined from the misreconstructed sample. For \Hel, the sum of symmetric 
Chebyshev polynomials up to second order is used. The variables \Dt\ and $q$ are modeled with the same PDF shape as the correctly reconstructed signal events but with an effective lifetime 
rather than $\tau_{B^{0}}$. This lifetime is obtained from MC and is necessary due to the presence of a tag-side track in the vertex reconstruction. This has the effect of reducing the average 
\Dz\ separation between \Brec\ and \Btag. We found from MC that, although the vertex reconstruction was incorrect, the $CP$ information was mostly retained, so the $CP$ parameters are shared with 
signal and are free in the fit to data. The difference between the generated $CP$ parameters in MC and misreconstructed signal events is then considered in the systematics.

\subsection{Continuum model}
\label{Continuum Model}
The parametrization of the continuum model is based on the sideband data; however, all the shape parameters of \De, \Mbc, \FD, and \mw\ are floated in the fit to data. \De\ and \Mbc\ are modeled by a linear and an ARGUS function, 
respectively, with parameters defined in bins of $s,d$. The variable \FD\ is modeled by the sum of either one or two Gaussian functions in each $l,s,d$ bin depending on the amount of data available in each bin. The PDF for \mw\ is a 
combination of a Gaussian and a linear function, while \Hel\ is the sum of a Gaussian centered around zero and a constant. The parameters of both these PDFs are determined in each $s,d$ bin. The \Dt\ model is fixed from the sideband,
\begin{equation}
  {\cal P}^{l,s,d}_{\qqbar}(\Dt, q) \equiv \frac{1}{2}\biggl[(1 - f^{d}_{\delta}) \frac{e^{-|\Dt|/\tau^{d}_{\qqbar}}}{2\tau^{d}_{\qqbar}} + f^{d}_{\delta} \; \delta( \Dt - \mu^{s,d}_{\delta})\biggr] \otimes R^{s,d}_{\qqbar}(\Dt),
\end{equation}
and contains a finite-lifetime and prompt component with a fraction $f^d_{\delta}$ and a mean $\mu^{d,s}_{\delta}$. The two components account for the long-lived charm and charmless contributions, respectively. The 
total distribution is convolved with a sum of two Gaussian functions
\begin{equation}
  R^{s,d}_{\qqbar}(\Dt) \equiv (1-f^{s,d}_{\rm tail})G(\Dt; \mu^{s,d}_{\rm mean}, S^{s,d}_{\rm main}\sigma) + f^{s,d}_{\rm tail}G(\Dt; \mu^{s,d}_{\rm mean}, S^{s,d}_{\rm main}S^{s,d}_{\rm tail}\sigma),
\end{equation}
which uses the event-dependent \Dt\ error constructed from the estimated vertex resolution $\sigma \equiv (\sqrt{\sigma^{2}_{\rm Rec}+\sigma^{2}_{\rm Tag}})/\beta \gamma c$ as a scale factor of 
the width parameters $S^{s}_{\rm main}$ and $S^{s}_{\rm tail}$.

\subsection{$\bm{\BBbar}$ model}
\label{BBbar Model}

The next-largest background comes from neutral and charged charm $b \rightarrow c$ and neutral and charged charmless $b \rightarrow u,d,s$ decays of the $B$ meson.
 Some of these $B$ decays exhibit peaking structure in the signal region due to the reconstruction of particular 
channels with identical final states. These are modeled separately from the nonpeaking \BBbar\ background, which is described in this subsection. The charm and charmless $B$ meson background shapes are determined from a large sample of MC simulation events based on $b \rightarrow c$ and $b \rightarrow u,d,s$ transitions, respectively.
The two data sets are further subdivided into neutral and charged $B$ meson samples to account for their different effective lifetimes.

For all \BBbar\ background shapes except for the charged charm samples, the \De\ distribution is modeled with the sum of a linear function and a Gaussian accounting for six-pion final states from which only 
five pions were reconstructed and thus peak roughly around $-0.14 \, {\rm GeV}/c^{2}$. The remaining charged charm samples are modeled with the sum of Chebyshev polynomials up to second order.
We model \Mbc\ in the neutral charm category with an ARGUS function and in the charged charm category with a histogram PDF. In the charmless models, the PDF for \Mbc\ is the sum of an 
asymmetric Gaussian and an ARGUS function. A sizable correlation of 12\% is found between \Mbc\ and \FD\ in the neutral charmless model for \BoKs\ and in the charged charmless model for \BpKp, 
which is taken into account by further parametrizing this shape of \Mbc\ in terms of \FD,
\begin{eqnarray}
 {\cal P}^{l,s,d}_{\rm Charmless}(\Mbc|\,\FD) &\equiv& f^{s,d}G(\Mbc; \, \mu^{s,d}, \, \sigma_{l}^{s,d}, \, \sigma_{r}^{s,d}) \nonumber \\ 
							&+& (1-f^{s,d})A(\Mbc; a^{s,d}+\gamma^{s,d} \FD, E_{\rm beam} ),
\label{e_rcmbc_wkp}
\end{eqnarray} 
where $\gamma^{s,d}$ is the correlation parameter. In \FD, the shape 
borrows from the signal model where, again, the main mean in each flavor-tag bin is obtained from the relevant \BBbar\ MC simulation sample. In the charm 
samples, \mw\ is modeled with a linear function; in the charmless samples, an additional Gaussian component is necessary. The variable \Hel\ in the charm sample is taken to be 
a histogram PDF; in the charmless model, the sum of a Gaussian and a linear function is used. We fit \Dt\ and $q$ with the same lifetime function as for the signal, 
but instead of the world average for the $B$ meson lifetime we determine an effective lifetime of the various $B$ meson decays from their respective MC samples. In general, the effective 
lifetime is smaller than the generated $B$ meson lifetime because a track in \Brec\ can originate from the tag side. The $CP$ parameters are fixed to zero.

\subsection{Peaking charm $\bm{\BBbar}$ model}
\label{Peaking Charm BBbar Model}
In the neutral decay mode \BoKs, this category includes the charm decays $B^{0}\rightarrow D^{*-}[\bar{D^{0}}\{\Ks\piz\}\pim]\pip$, 
$B^{0} \rightarrow  D^{-}[\Ks \pim \piz]\pip$ and $B^{0} \rightarrow D^{-}[\Ks \pim]\rho^{+}[\pip \piz]$. For the charged decay mode \BpKp, this includes only $B^{+} \rightarrow D^{0}[K^{+}\pim] \rho^{+}[\pip \piz]$. 
To account for the peaking structure in \De, its PDF is taken to be the same as that of the signal; however, the parameters of the linear component and its relative fraction are determined from the 
peaking charm \BBbar\ MC simulation. The model for \Mbc\ is taken to be the combination of a Gaussian function and an ARGUS function. Because of a correlation between \De\ and 
\Mbc\ of up to 12\%, the fraction of the Gaussian component is linearly parametrized in terms of \De.  The variable \FD\ also borrows from the signal model with the main mean and width of the distribution 
in each flavor-tag bin determined from the peaking charm \BBbar\ MC simulation. In the neutral mode, \Hel\ is modeled with a Gaussian centered around zero; in the charged mode, a symmetrized 
histogram is used. The variables \Dt\ and $q$ are fitted with a lifetime function with an effective lifetime determined from MC simulation.

\subsection{Peaking charmless $\bm{\BBbar}$ model}
\label{Peaking Charmless BBbar Model}
This category only affects the charged decay mode \BpKp, and includes the charmless decays 
$B^{+} \rightarrow {a_{1}^{0}}[\pip\pim\piz] K^{+}$ and $B^{+} \rightarrow {\omega}[\pip\pim\piz] \pi^{+}$.
In \De, two peaks are visible in the distribution: one around zero and one shifted to positive values near the difference between the kaon and pion mass, originating from 
$B^{+} \rightarrow \omega[\pip\pim\piz] \pip$ decays, where a pion is misidentified as a kaon. Both peaks are modeled with the triple Gaussian component of the signal PDF for \De, where the 
mean of the misidentified peak is determined from the peaking charmless \BBbar\ MC simulation. The combinatorial component is modelled with a first-order Chebyshev, for which the shape and fraction are 
also determined from MC. The model for \Mbc\ is the sum  of two Gaussians and an ARGUS function. Because of up to a 14\% 
correlation between \Mbc\ and \De, 
the main width and fraction of  the Gaussian as well as the ARGUS slope parameter are parametrized in terms of \De. Once again, \FD\ borrows from the signal model with the main mean and 
width in each flavor-tag bin determined from peaking charmless \BBbar\ MC simulation. The \Hel\ variable is modeled with the sum of symmetric Chebyshev polynomials up to fourth order. Finally, 
\Dt\ and $q$ are fitted with a lifetime function with an effective lifetime determined from MC simulation.

\subsection{Full model}
\label{Full Model}
The total extended likelihood is given by
\begin{equation}
  {\cal L} \equiv \prod_{l,s,d} \frac{e^{-\sum_{j}N^{s,d}_{j}f^{l,s,d}_{j}}}{N_{l,s,d}!} \prod^{N_{l,s,d}}_{i=1} 
  \sum_{j}N^{s,d}_{j}f^{l,s,d}_{j} {\cal P}^{l,s,d}_{j}(\De^{i}, \Mbc^{i}, \FD^{i}, \mw^{i}, \Hel^{i}, \Dt^{i}, q^{i}),
\end{equation}
which iterates over $i$ events, $j$ categories, $l$ flavor-tag bins, $s$ detector configurations, and the $d$ data samples \BoKs\ and \BpKp. The fraction of events in each $l,s,d$ bin, for category $j$, 
is denoted by $f^{l,s,d}_{j}$. With the exception of continuum, for which the fit fractions are free in the fit to the data, these parameters are fixed for each category from their respective MC samples. The 
fraction of signal events in each $l,s,d$ bin, $f^{l,s,d}_{\rm Sig}$, is corrected using common correction factors for \BoKs\ and \BpKp, which are also free parameters in the fit to the data. 
Additional free fit parameters include the 
$N^{s,d}_{\qqbar}$ yields. Instead of obtaining separate signal yields for SVD1 and SVD2 $N^{s,d}_{\rm Sig}$, this parameter is transformed so that the branching fraction becomes a single free parameter 
between $s$ samples and is incorporated into the fit with
\begin{equation}
 N^{s,d}_{\rm Sig} = {\cal B}^d(\BK) \times N_{\BBbar}^{s} \epsilon_{\rm Sig}^{s,d} \eta^{s,d}_{\rm Sig},
\end{equation}
where $N_{\BBbar}^{s}$ is the number of \BBbar\ pairs collected by the Belle detector and $\epsilon_{\rm Sig}^{s,d}$ and $\eta^{s,d}_{\rm Sig}$ are detection efficiencies and selection criteria correction factors, 
respectively, given in Table~\ref{tab_eff_pid}. The yield of the misreconstructed signal events is fixed with respect to the signal yield with a relative fraction determined from MC. The remaining $N^{s}_{\BBbar}$ 
yields are fixed from their expected amounts as determined from MC simulation and given in Table~\ref{tab_bf_fixed}. In total, there are  204 free parameters in the fit: 54 belonging to signal and the remaining 150 to 
the continuum.

Following this, two additional fits are performed to calculate \Acpc\ by measuring the two terms in Eq. (\ref{eq_awk}). In these fits, we divide the \BpKp\ sample based on the kaon charge so that we may separately 
extract ${\cal B}(\Bp \to \omega \Kp)$ and ${\cal B}(B^- \to \omega \Km)$. The only three free parameters in these fits are the signal branching fraction and the continuum yields $N_{\qqbar}^{s,w\Kp}$. The \BBbar\ 
yields for each \BpKp\ subsample are also recalculated as needed based on the relevant kaon charge. All remaining parameters are fixed to those found in the initial fit to the data.

\begin{table}
  \centering
  \caption{Summary of yields fixed relative to other yields in the fit for \BoKs\ (top) and \BpKp\ (bottom). The values of the yields and their uncertainties are obtained from MC statistics.}
  \begin{tabular}
    {@{\hspace{0.5cm}}l@{\hspace{0.5cm}} @{\hspace{0.5cm}}l@{\hspace{0.5cm}}  @{\hspace{0.5cm}}l@{\hspace{0.5cm}}}
    \hline \hline 
    Yield & SVD1 & SVD2\\
    \hline
    $N^{s,\omega\Ks}_{\rm Mis}$ & $(0.0192 \pm 0.0004)N^{1,\omega\Ks}_{\rm Sig}$ & $(0.0187 \pm 0.0004)N^{2,\omega\Ks}_{\rm Sig}$\\
    $N^{s,\omega\Ks}_{{\rm Charm} \; \BzBzb}$ & $12 \pm 3$ & $56 \pm 7$\\
    $N^{s,\omega\Ks}_{{\rm Charm} \; \BpBm}$ & $(1.066 \pm 0.094)N^{1,\omega\Ks}_{{\rm Charm} \; \BzBzb}$ & $(1.268 \pm 0.048)N^{2,\omega\Ks}_{{\rm Charm} \; \BzBzb}$\\
    $N^{s,\omega\Ks}_{{\rm Charmless} \; \BzBzb}$ & $(5.992 \pm 0.216)N^{1,\omega\Ks}_{{\rm Charm} \; \BzBzb}$ & $(7.191 \pm 0.109)N^{2,\omega\Ks}_{{\rm Charm} \; \BzBzb}$\\
    $N^{s,\omega\Ks}_{{\rm Charmless} \; \BpBm}$ & $(4.537 \pm 0.193)N^{1,\omega\Ks}_{{\rm Charm} \; \BzBzb}$ & $(6.295 \pm 0.106)N^{2,\omega\Ks}_{{\rm Charm} \; \BzBzb}$\\
    $N^{s,\omega\Ks}_{{\rm Peaking \; Charm} \; \BzBzb}$ & $(0.719 \pm 0.077)N^{1,\omega\Ks}_{{\rm Charm} \; \BzBzb}$ & $(0.780\pm 0.036)N^{2,\omega\Ks}_{{\rm Charm} \; \BzBzb}$\\
    \hline \hline
  \end{tabular}

 \vspace{0.5cm}

  \begin{tabular}
    {@{\hspace{0.5cm}}l@{\hspace{0.5cm}} @{\hspace{0.5cm}}l@{\hspace{0.5cm}}  @{\hspace{0.5cm}}l@{\hspace{0.5cm}}}
    \hline \hline 
    Yield & SVD1 & SVD2\\
    \hline
    $N^{s,\omega\Kp}_{\rm Mis}$ & $(0.0182 \pm 0.0003)N^{1,\omega\Kp}_{\rm Sig}$ & $(0.0182 \pm 0.0003)N^{2,\omega\Kp}_{\rm Sig}$\\
    $N^{s,\omega\Kp}_{{\rm Charm} \; \BzBzb}$ & $25 \pm 5$ & $147 \pm 12$\\
    $N^{s,\omega\Kp}_{{\rm Charm} \; \BpBm}$ & $(3.334 \pm 0.115)N^{1,\omega\Kp}_{{\rm Charm} \; \BzBzb}$ & $(2.808 \pm 0.044)N^{2,\omega\Kp}_{{\rm Charm} \; \BzBzb}$\\
    $N^{s,\omega\Kp}_{{\rm Charmless} \; \BzBzb}$ & $(6.000 \pm 0.147)N^{1,\omega\Kp}_{{\rm Charm} \; \BzBzb}$ & $(5.556 \pm 0.060)N^{2,\omega\Kp}_{{\rm Charm} \; \BzBzb}$\\
    $N^{s,\omega\Kp}_{{\rm Charmless} \; \BpBm}$ & $(9.913 \pm 0.198)N^{1,\omega\Kp}_{{\rm Charm} \; \BzBzb}$ & $(8.828 \pm 0.077)N^{2,\omega\Kp}_{{\rm Charm} \; \BzBzb}$\\
    $N^{s,\omega\Kp}_{{\rm Peaking \; Charm} \; \BpBm}$ & $(1.504 \pm 0.077)N^{1,\omega\Kp}_{{\rm Charm} \; \BzBzb}$ & $(1.300 \pm 0.029)N^{2,\omega\Kp}_{{\rm Charm} \; \BzBzb}$\\
    $N^{s,\omega\Kp}_{{\rm Peaking \; Charmless} \; \BpBm}$ & $(7.130 \pm 0.168)N^{1,\omega\Kp}_{{\rm Charm} \; \BzBzb}$ & $(6.792 \pm 0.068)N^{2,\omega\Kp}_{{\rm Charm} \; \BzBzb}$\\
    \hline \hline
  \end{tabular}
  \label{tab_bf_fixed}
\end{table}

\subsection{Fit validation and improvements}

To determine the branching fractions and $CP$ violation parameters, in contrast to the previous Belle analyses~\cite{wks_Belle,wk_Belle}, we fit all variables and the two decay channels 
simultaneously. 
Extracting common calibration parameters between the two decay modes from the data allows us to neglect systematic uncertainties in the low statistics neutral mode arising 
from the difference between data and MC simulation.
An important difference from the previous Belle analyses is the improved tracking algorithm applied to the SVD2 data sample. This, combined with a looser cut on 
${\cal{L}}_{\BBbar/\qqbar}$, improves the efficiency compared to the previous Belle branching fraction analysis~\cite{wk_Belle} by a factor of 4 for the neutral mode and 2 for the charged mode. 
To improve the statistical precision of the branching fraction over the previous measurement, \FD\ has been included in the fit. Another improvement over both previous analyses is the inclusion of 
the \Hel\ observable into the fit, which significantly improves background discrimination. To determine the $CP$ parameters, the previous Belle analysis~\cite{wks_Belle} applied a two-step procedure 
where an initial fit without \Dt\ and $q$ was performed to obtain a signal yield. This allowed the event-dependent probabilities of each component to be determined and then used as input to set the 
fractions of each component in a fit to \Dt\ and $q$. Our procedure of combining all variables together in a single fit has the added benefit of further discrimination against the continuum with the 
\Dt\ variable and makes the treatment of systematic uncertainties more straightforward, at a cost of analysis complexity and longer computational time.

To test the validity of this model, we determine a possible fit bias from a pseudoexperiment MC simulation study in which the signal and the \BBbar\ backgrounds are generated from GEANT-simulated 
events while the continuum background is generated from our model of the sideband data. We find a bias for the branching fraction values of 16\% and 45\% of their statistical uncertainties for the 
neutral and charged mode, respectively. We correct the central values by these amounts and assign half the bias as a systematic uncertainty. Additionally, a linearity test across the physical 
\Acpm-\Scpm\ region is performed, showing no significant bias. This pseudoexperiment study indicates 30\% improvement in the statistical uncertainty of the branching fractions of \BK, 15-20\% 
improvement in the statistical uncertainty of the \BoKs\ time-dependent $CP$ parameters and 30\% improvement of \Acpc\ over the previous analysis methods. These numbers are calculated by 
scaling all uncertainties from the previous analyses to that expected with the final data set.

To test the validity of the \Dt\ resolution description and reconstruction procedure, we perform a separate fit releasing the \Bz\ and \Bp\ lifetimes while blinding the physics parameters; 
the results for $\tau_{\Bz}$ and $\tau_{\Bp}$ are consistent with their respective current world averages~\cite{PDG} within two standard deviations. As a further check of the \Dt\ resolution 
function and the parameters describing the probability of mistagging, we fit for the time-dependent $CP$ parameters of the \BpKp\ sample by substituting Eq. (\ref{eq:dtq}) for Eq. (\ref{eq:dtq_wk}); 
the results are consistent within one standard deviation with \Acpc\ obtained from the nominal fit and with null asymmetry for \Scpc.

\section{Results}
\label{Results}
\begin{figure}
  \centering
  \includegraphics[height=180pt,width=!]{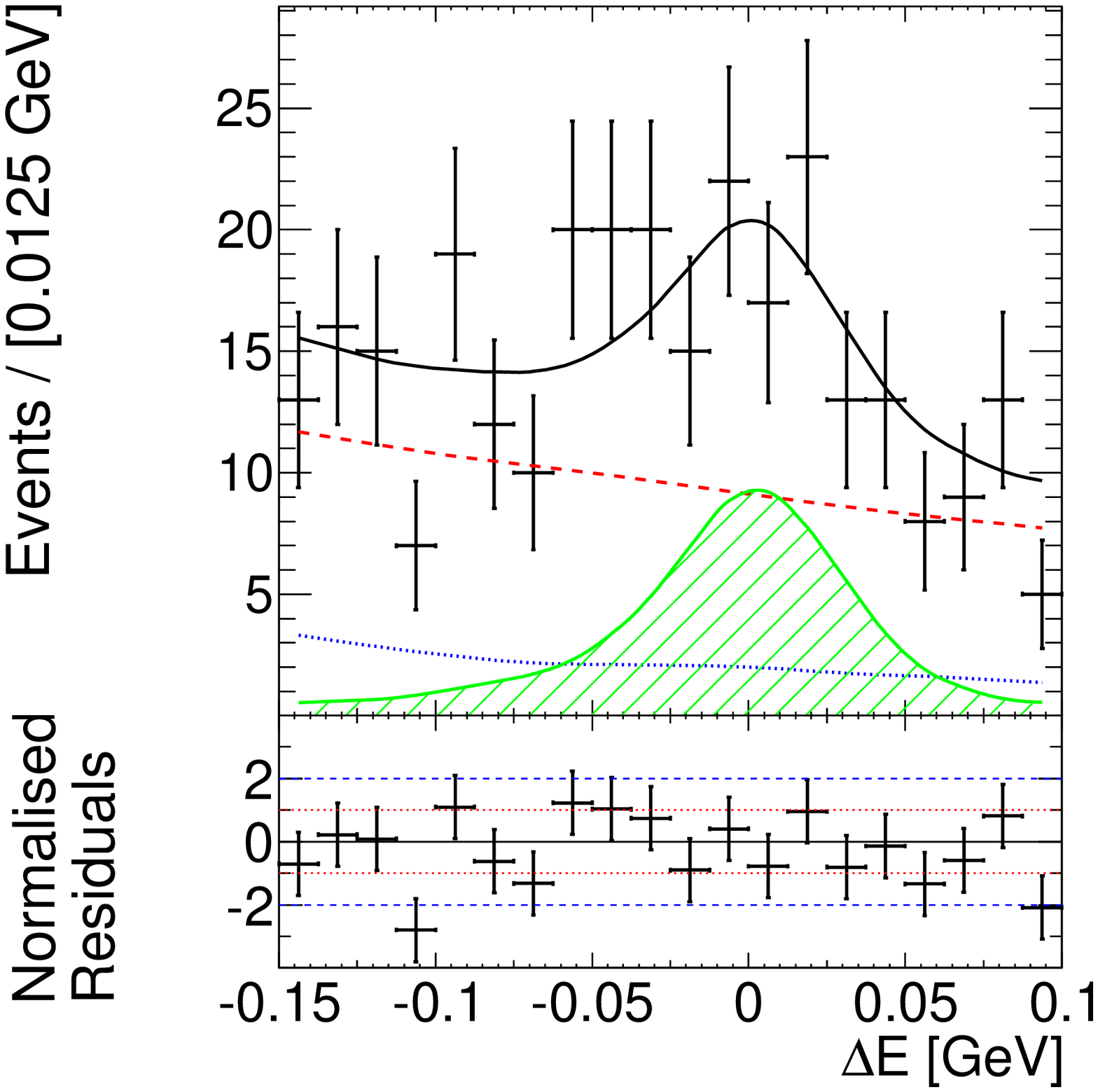}
  \includegraphics[height=180pt,width=!]{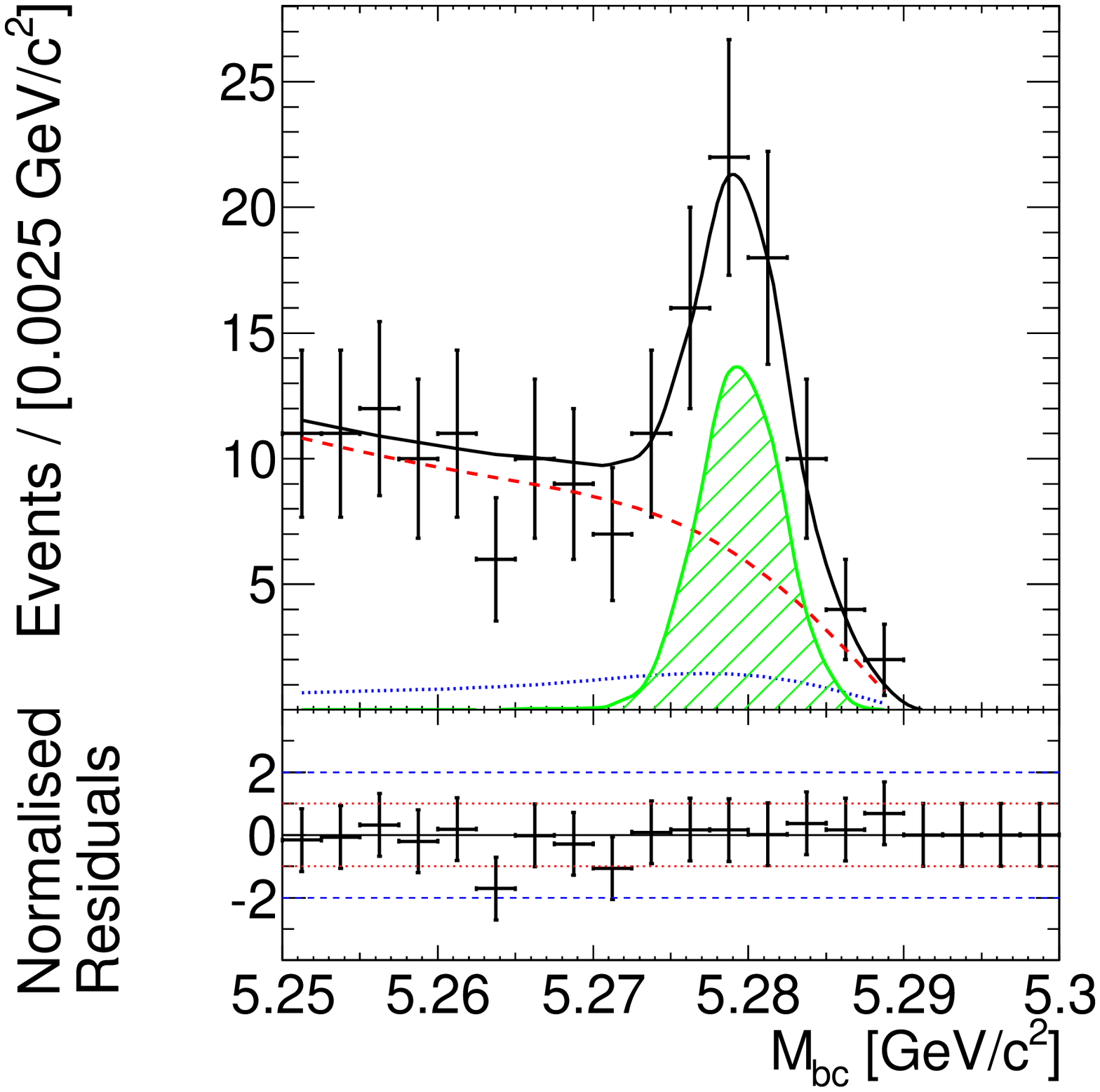}
  \put(-227,155){(a)}
  \put(-40,155){(b)}

  \includegraphics[height=180pt,width=!]{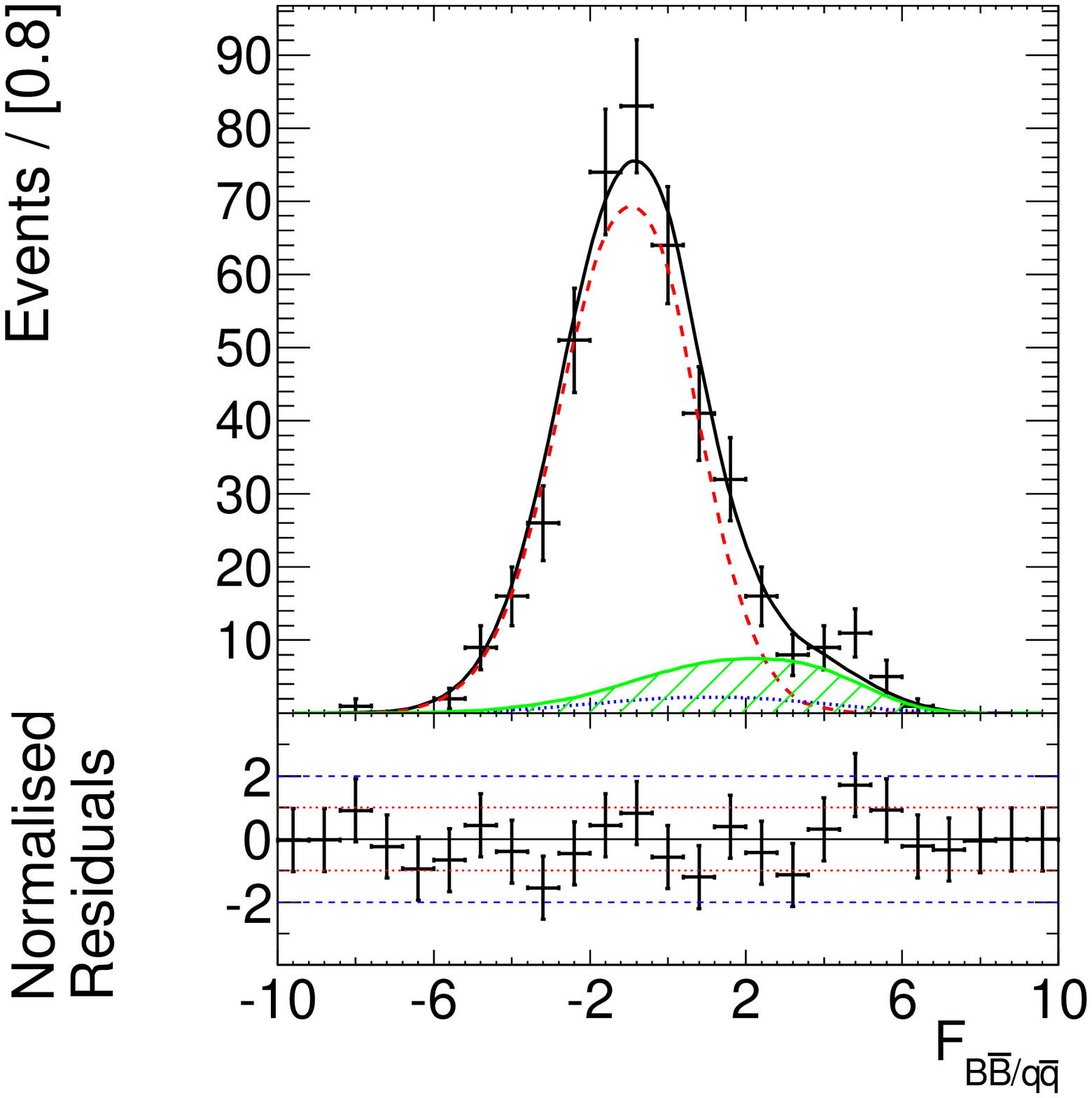}
  \includegraphics[height=180pt,width=!]{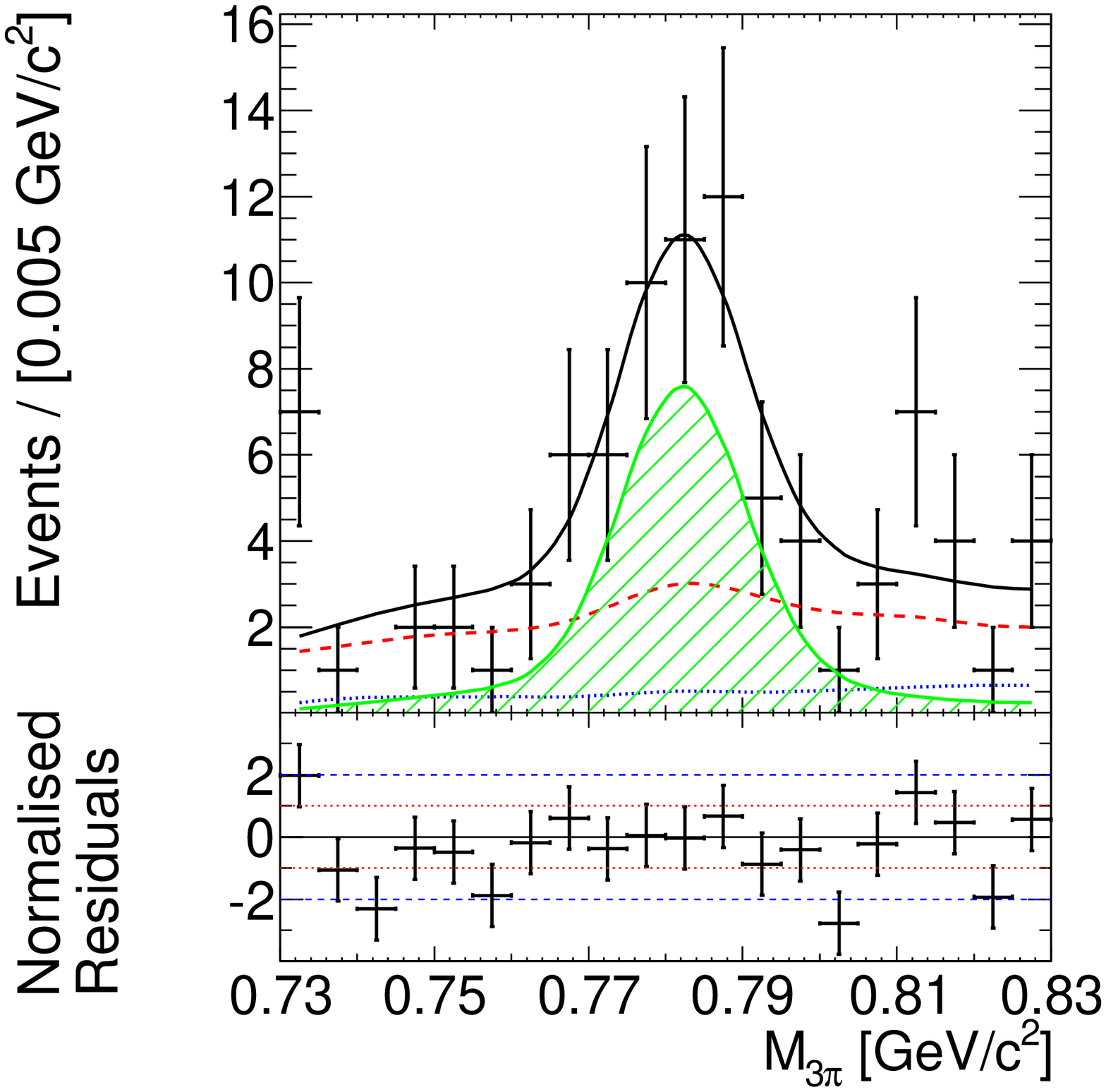}
  \put(-227,155){(c)}
  \put(-40,155){(d)}

  \includegraphics[height=180pt,width=!]{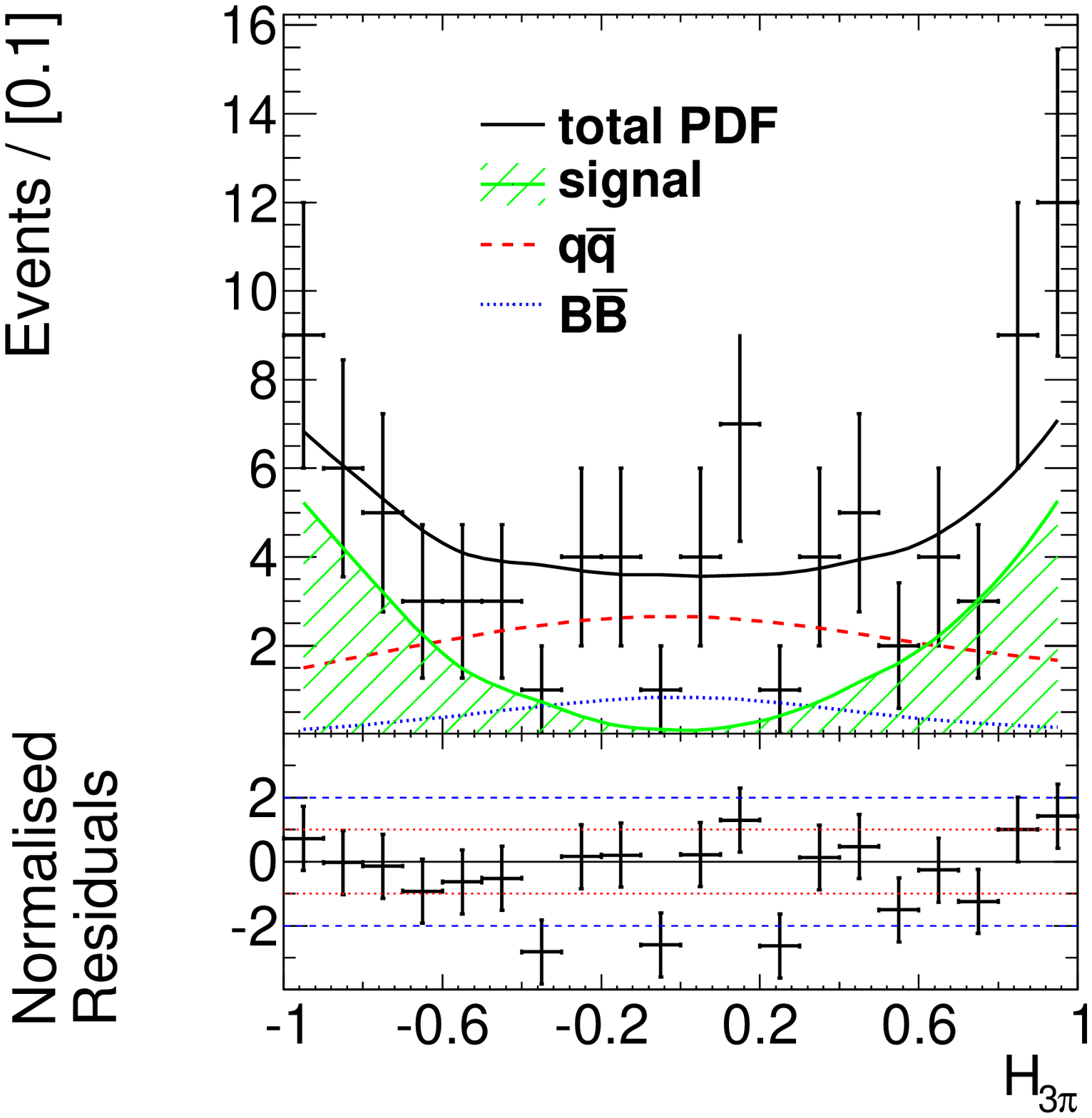}
  \put(-40,155){(e)}

  \caption{Projections of the fit to the  \BoKs\ data enhanced in the signal region. Points with error bars represent the data, and the solid black curves or histograms represent 
  the fit results. The signal enhancements, $ -0.04 \textrm{ GeV} < \De < 0.03 \textrm{ GeV}$, $\Mbc > 5.27 \textrm{ GeV}/c^2$, $\FD > 1$, and $r > 0.5$, except for the enhancement of the fit
  observable being plotted, are applied to each projection. (a), (b), (c), (d), and (e) show the \De, \Mbc, \FD, \mw, and \Hel\ projections, respectively. Green hatched curves show the \BoKs\ signal 
  component, dashed red curves indicate the \qqbar\ background, and blue dotted curves show the \BBbar\ background component.}
  \label{fig_data_wks}
\end{figure}

\begin{figure}
  \centering
  \includegraphics[height=180pt,width=!]{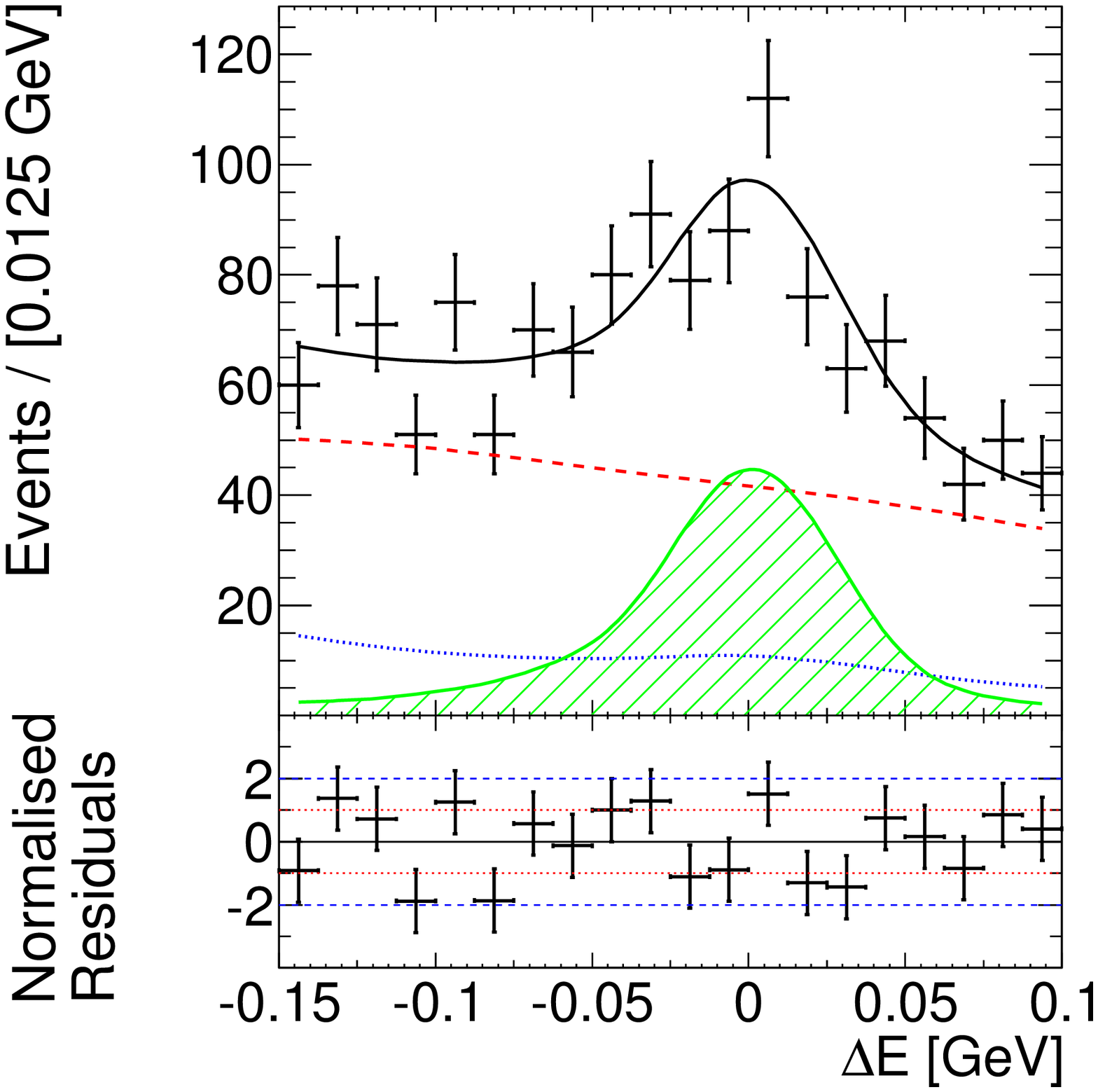}
  \includegraphics[height=180pt,width=!]{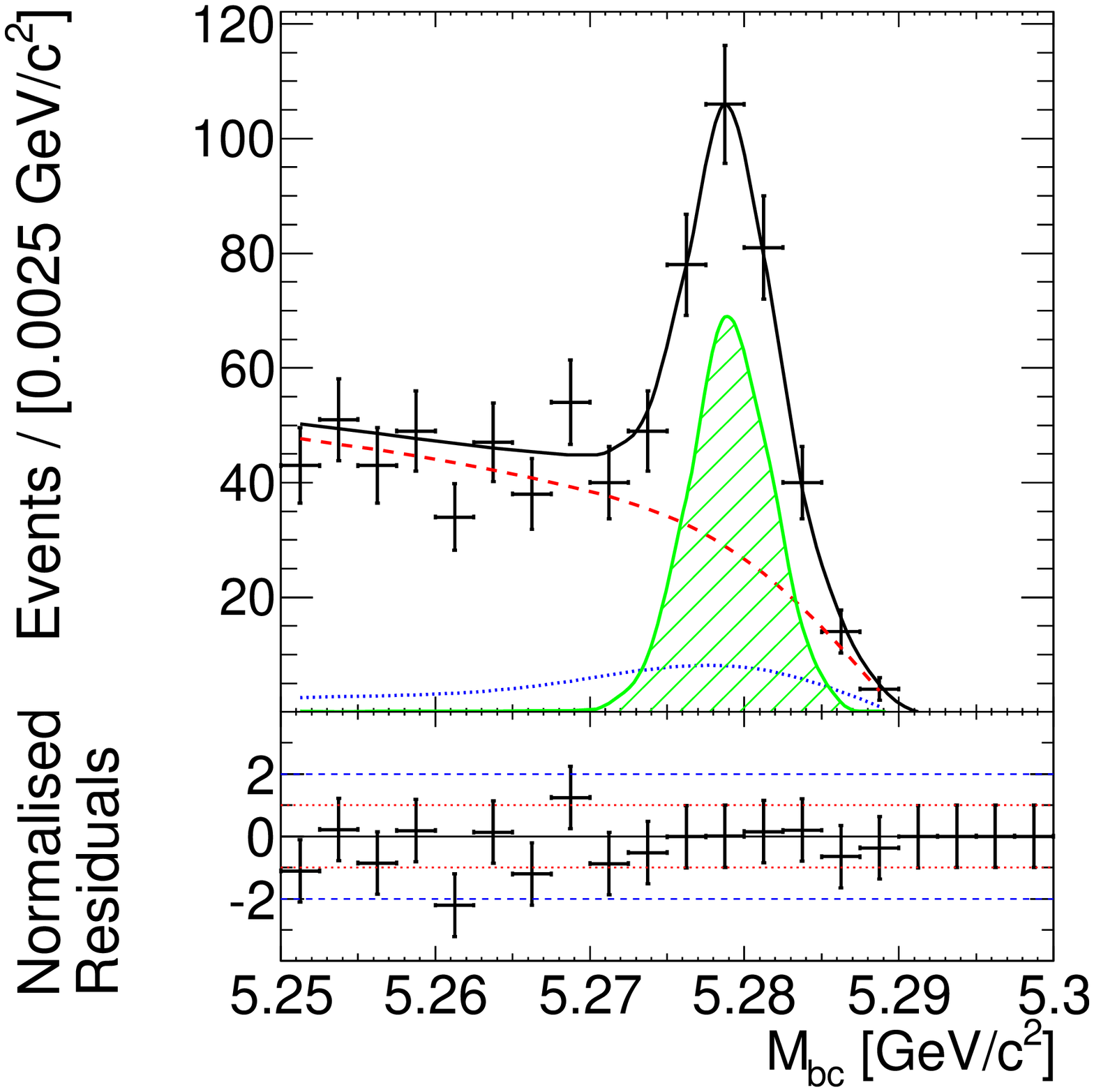}
  \put(-227,155){(a)}
  \put(-40,155){(b)}

  \includegraphics[height=180pt,width=!]{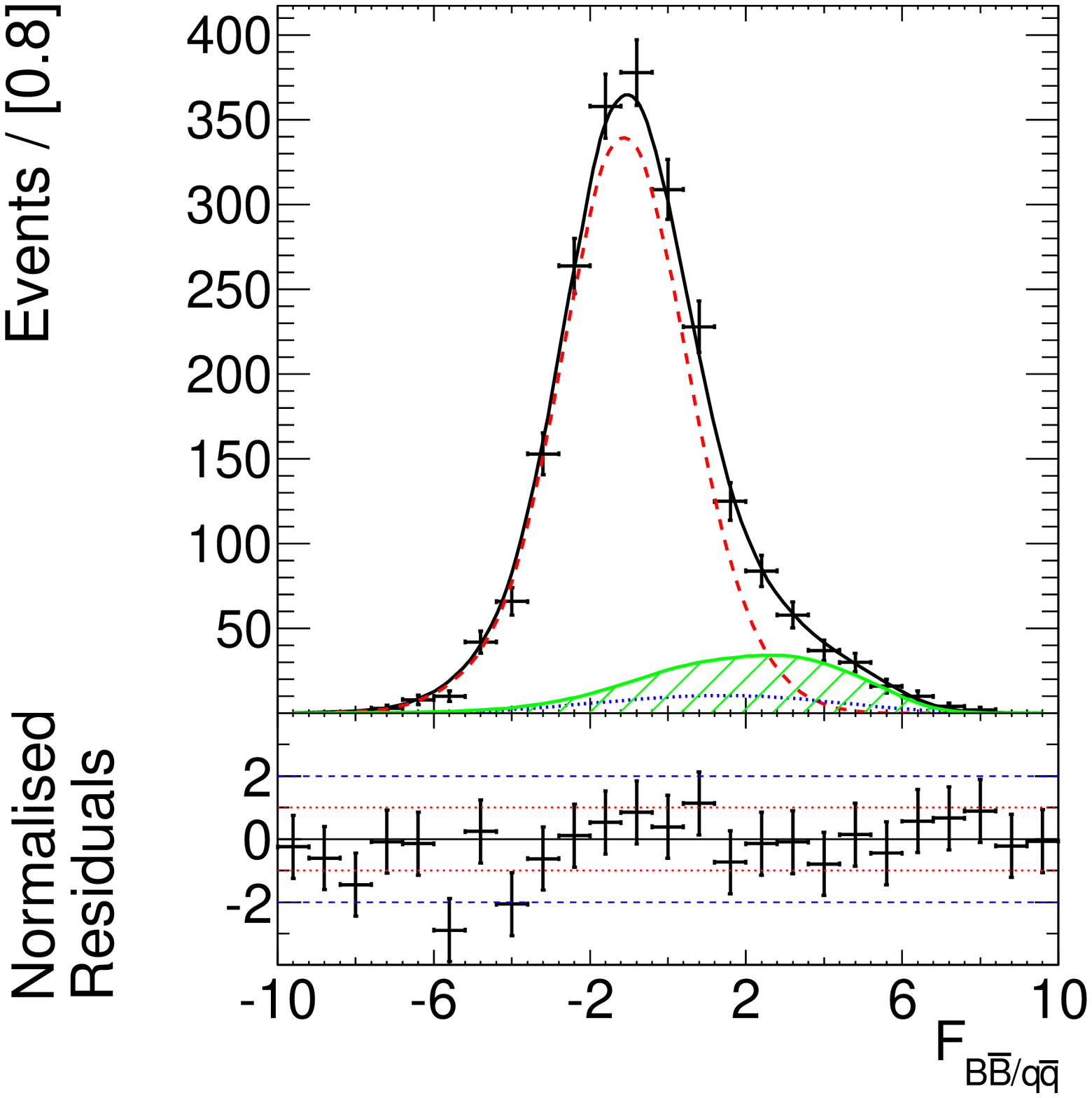}
  \includegraphics[height=180pt,width=!]{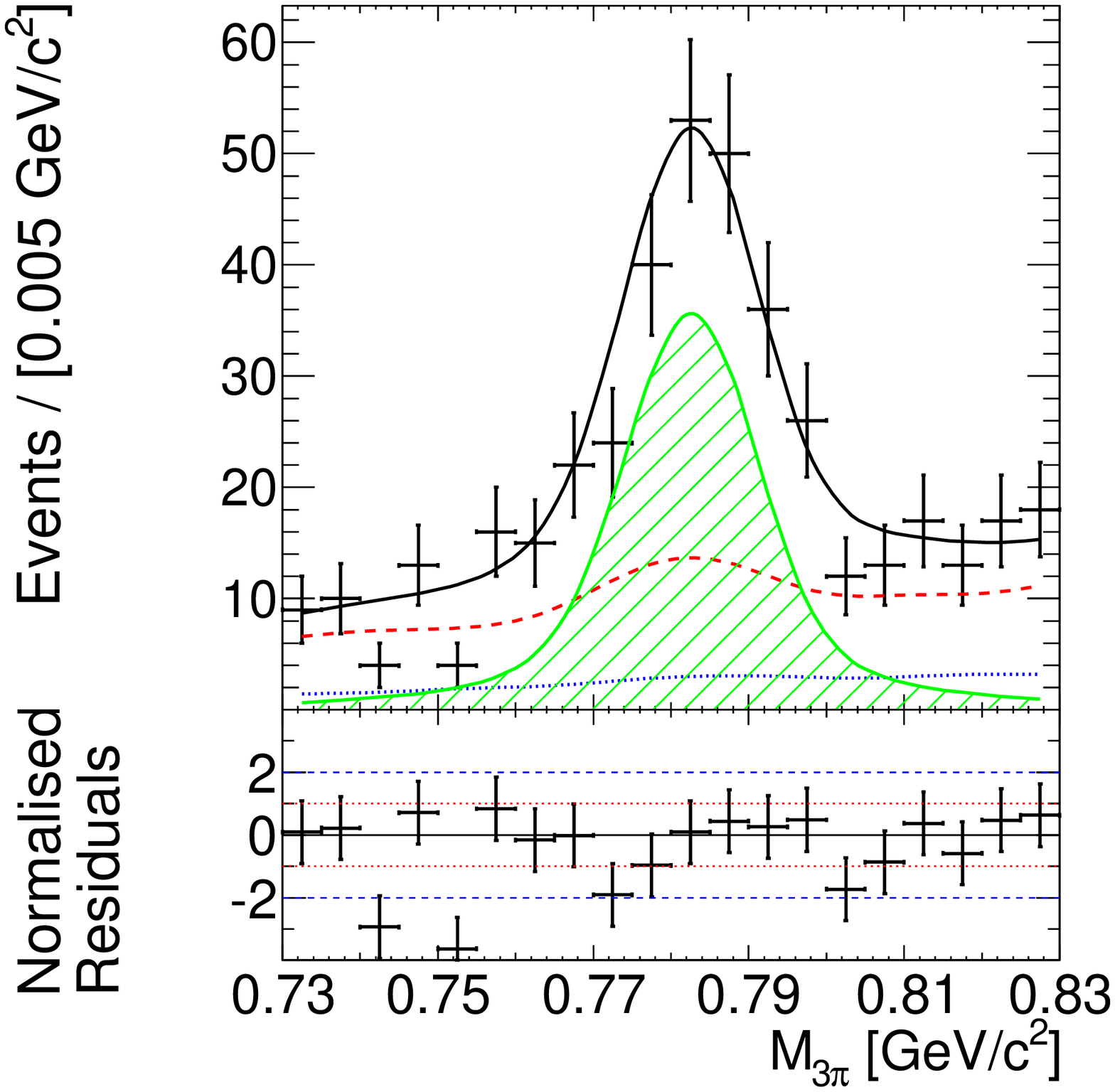}
  \put(-227,155){(c)}
  \put(-40,155){(d)}

  \includegraphics[height=180pt,width=!]{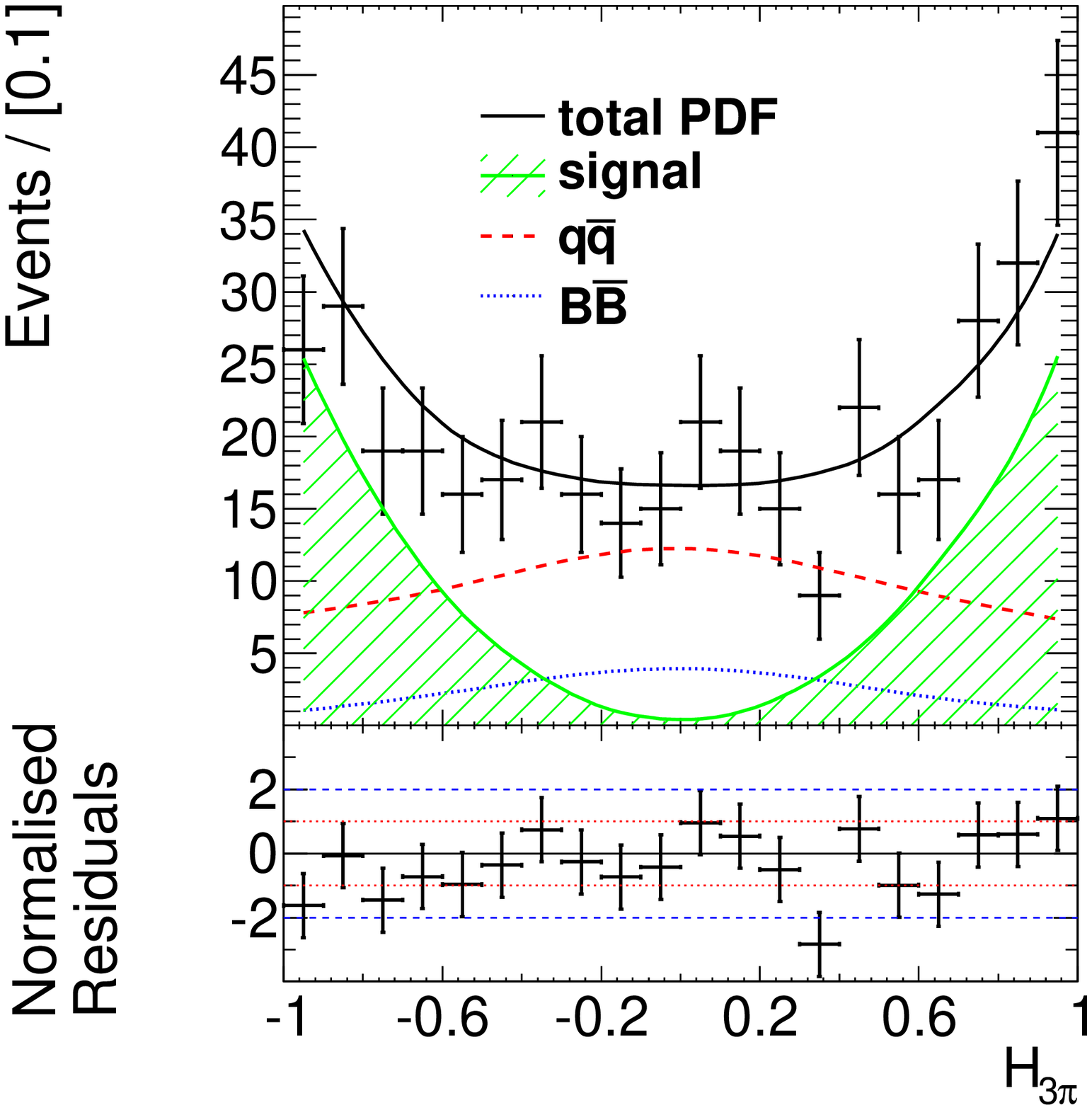}
  \includegraphics[height=180pt,width=!]{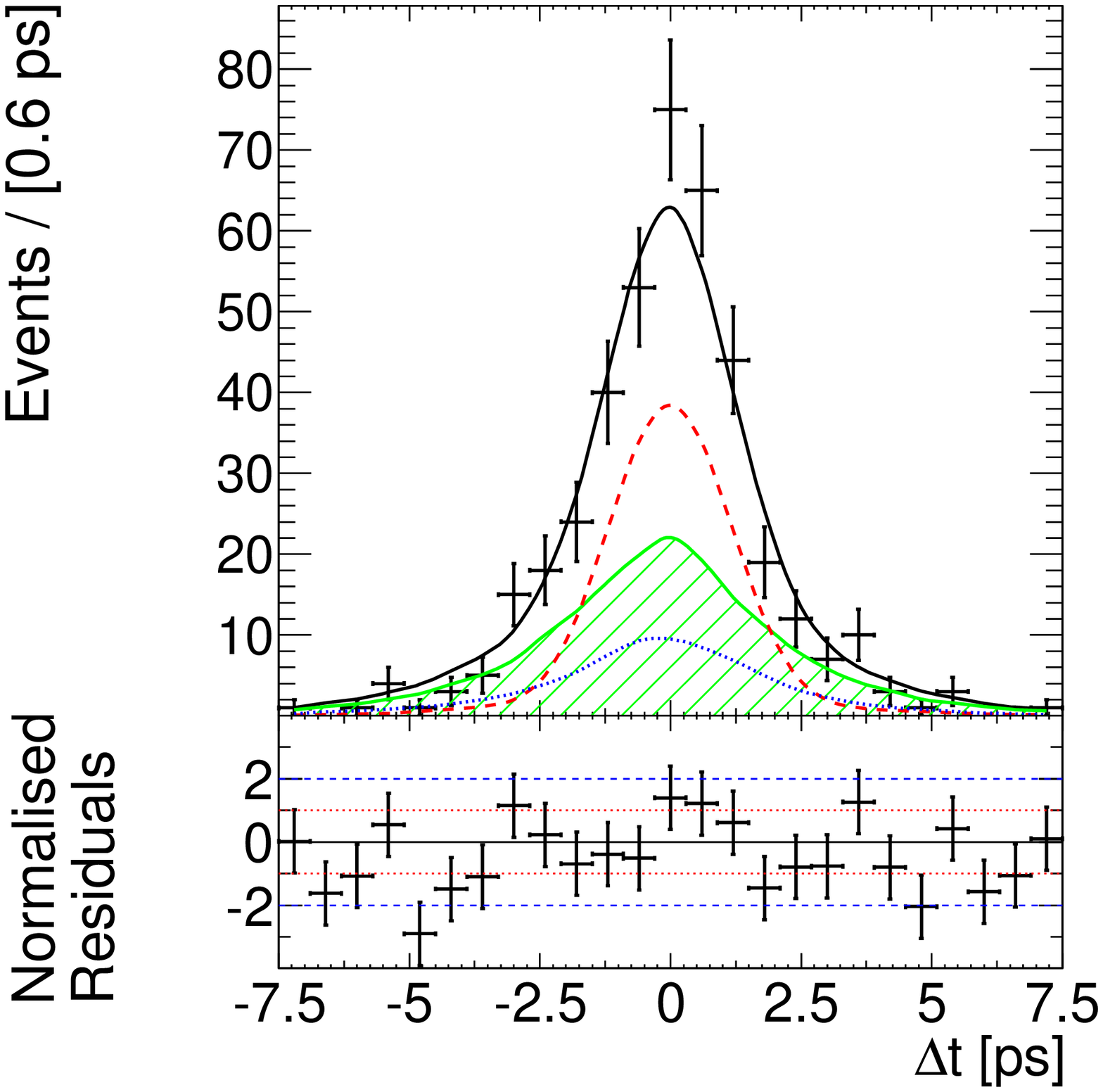}
  \put(-227,155){(e)}
  \put(-40,155){(f)}

  \caption{Projections of the fit to the \BpKp\ data enhanced in the signal region. Points with error bars represent the data, and the solid black curves or histograms represent 
  the fit results. The signal enhancements, $ -0.04 \textrm{ GeV} < \De < 0.03 \textrm{ GeV}$, $\Mbc > 5.27 \textrm{ GeV}/c^2$, $\FD > 1$, and $r > 0.5$, except for the enhancement of the fit
  observable being plotted, are applied to each projection. (a), (b), (c), (d), (e), and (f) show the \De, \Mbc, \FD, \mw, \Hel, and \Dt\ projections, respectively. Green hatched curves show the \BpKp\ 
  signal component, dashed red curves indicate the \qqbar\ background, and blue dotted curves show the \BBbar\ background component.}
  \label{fig_data_wkp}
\end{figure}

\begin{figure}
  \centering
  \includegraphics[height=220pt,width=!]{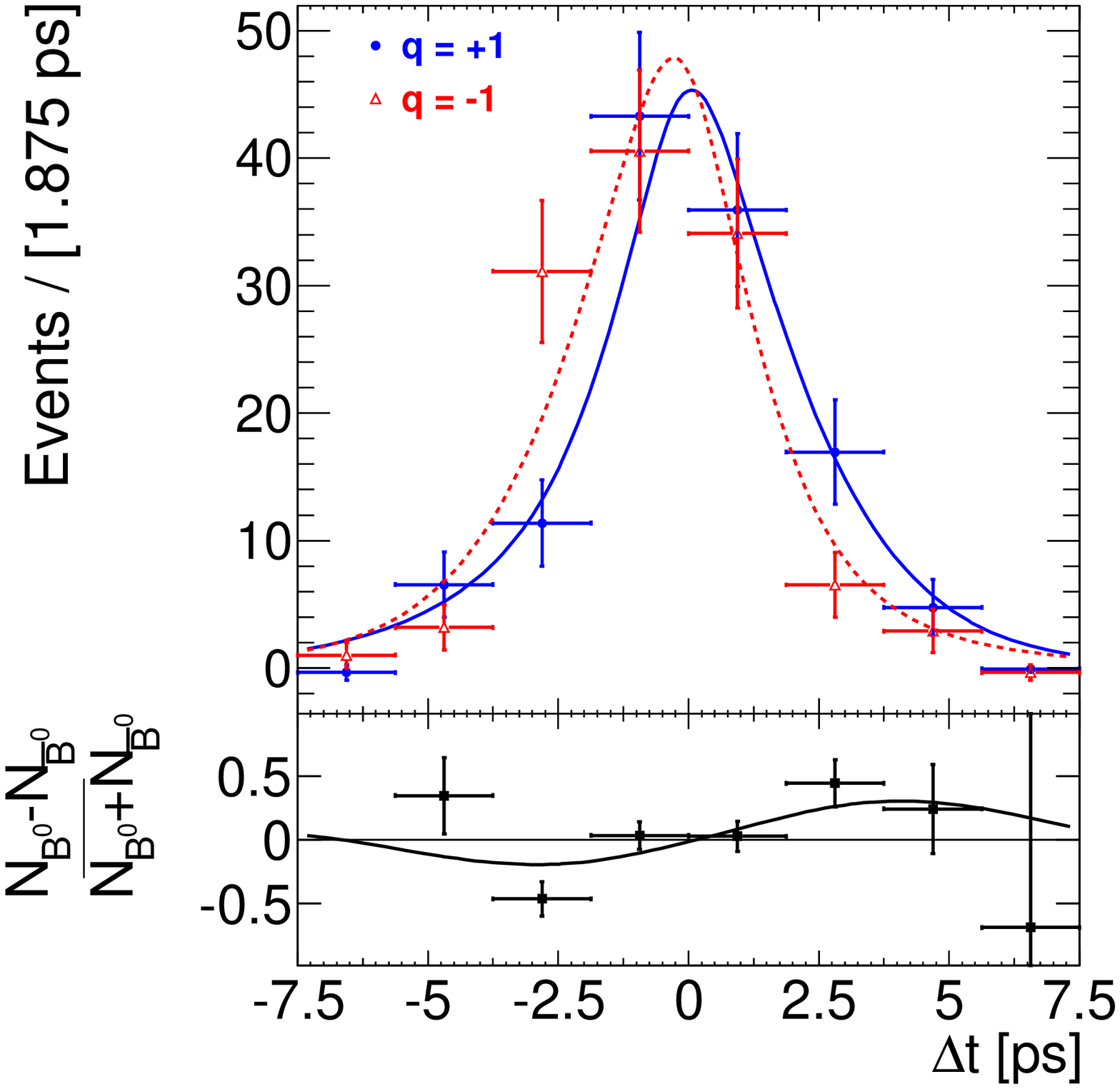}
  \put(-47,190){(a)}
  \put(-47,60){(b)}

  \caption{Background subtracted time-dependent fit results for \BoKs. (a) shows the \Dt\ distribution for each \Btag\ flavor $q$. The solid blue and dashed red curves represent the 
  \Dt\ distributions for \Bz\ and \Bzb\ tags, respectively. (b) shows the asymmetry of the plot above them, $(N_{\Bz} - N_{\Bzb})/(N_{\Bz} + N_{\Bzb})$, where $N_{\Bz}$ ($N_{\Bzb}$) is the measured 
  signal yield of \Bz\ (\Bzb) events in each bin of \Dt.}
  \label{fig_data_dt}
\end{figure}

From the fits to the data containing 17860 \BoKs\ and 88007 \BpKp\ candidates, the branching fractions and $CP$ violation parameters
  \begin{eqnarray}
    {\cal B}(\BoK) \!\!&=&\!\! (4.5 \pm 0.4 \textrm{ (stat)} \pm 0.3 \textrm{ (syst)})\times 10^{-6},\nonumber \\
    {\cal B}(\BpKp) \!\!&=&\!\! (6.8 \pm 0.4 \textrm{ (stat)} \pm 0.4 \textrm{ (syst)})\times 10^{-6},\nonumber \\
    \Acpm \!\!&=&\!\! -0.36 \pm 0.19 \textrm{ (stat)} \pm 0.05 \textrm{ (syst)},\nonumber \\
    \Scpm \!\!&=&\!\! +0.91 \pm 0.32 \textrm{ (stat)} \pm 0.05 \textrm{ (syst)},\nonumber \\
    \Acpc \!\!&=&\!\! -0.03 \pm 0.04 \textrm{ (stat)} \pm 0.01  \textrm{ (syst)},
  \end{eqnarray}
are obtained, where the first uncertainty is statistical and the second is systematic, which is discussed below (Sec.~\ref{Systematic Uncertainties}). 
The statistical correlation coefficients between the branching fractions and the $CP$ parameters are below $10^{-5}$ except for the 0.4\% correlation between \Acpm\ and \Scpm. 
Signal-enhanced fit projections are shown in Figs.~\ref{fig_data_wks}, \ref{fig_data_wkp}, and \ref{fig_data_dt}. The \BoKs\ and \BpKp\ branching fractions have been bias corrected, 
corresponding to signal event yields of $N(\BoKs) = 234 \pm 22$ and $N(\BpKp) = 1114 \pm 59$ where the uncertainties are 
statistical only. Before the bias correction, the central values of the branching fractions are ${\cal B}(\BoK) = 4.5\times 10^{-6}$ and ${\cal B}(\BpKp) = 6.9\times 10^{-6}$. From the yields 
obtained in the fit to the data, the relative contributions of each component in the neutral mode are found to be $1.3\%$ for the signal \BoKs, $96.5\%$  for the continuum, and $2.2\%$ for the \BBbar\ background. 
For the charged mode, we obtain $1.3\%$ signal \BpKp, $96.8\%$ continuum, and $1.9\%$ \BBbar. All results are consistent with the previous Belle measurements~\cite{wk_Belle,wks_Belle} within two 
standard deviations. The statistical errors obtained in our fit to the data agree with those expected obtained from the pseudoexperiment study mentioned in the previous section.

\begin{figure}
  \centering
  \includegraphics[height=200pt,width=!]{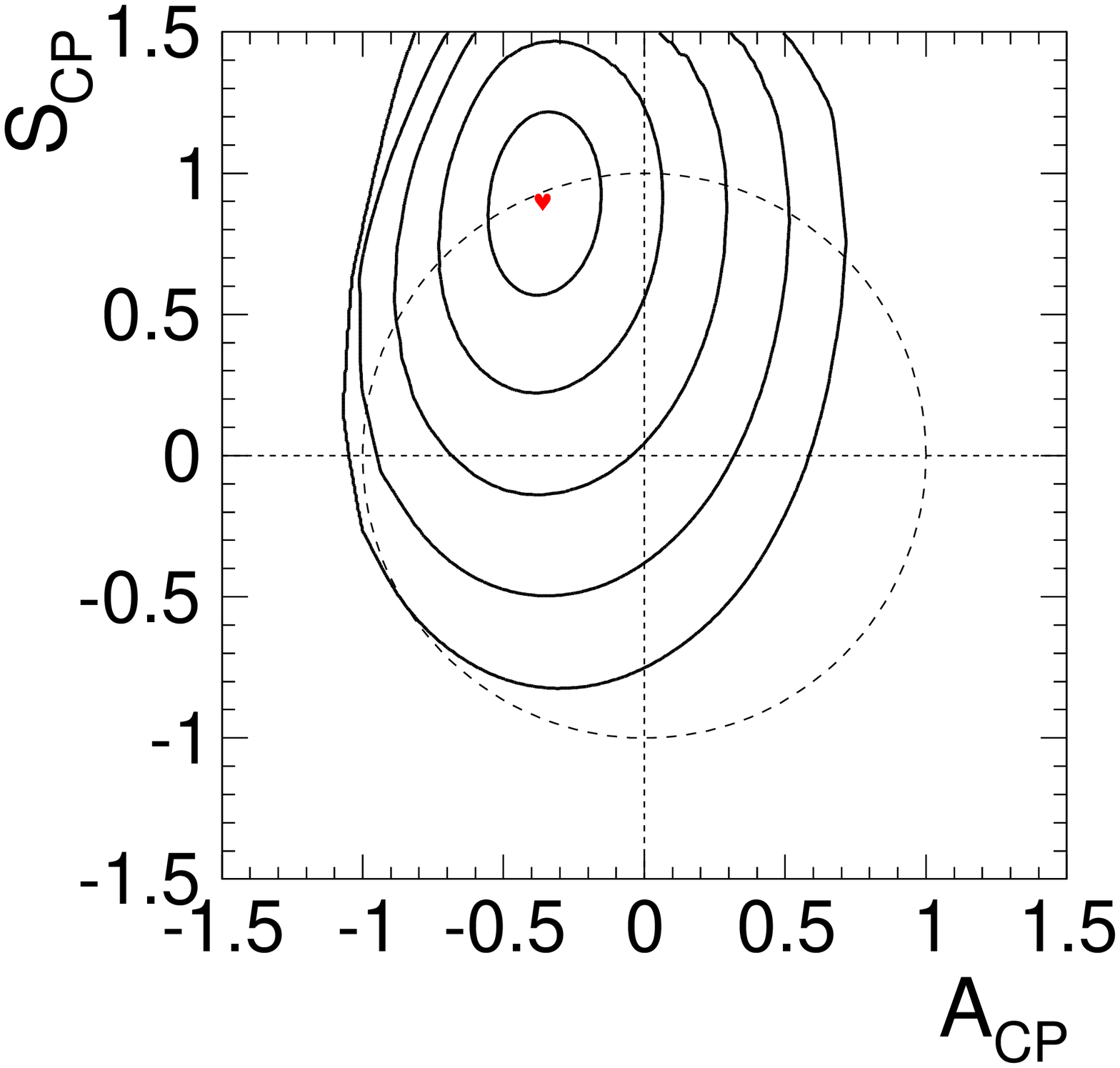}
  \caption{Likelihood scan in the \Acpm-\Scpm\ plane including systematic uncertainties. The dashed circle represents the physical boundary of $CP$ violation. Starting from the red marker in the 
  center that identifies the fit result, the concentric curves represent the contours from 1 to 5 standard deviations from the fit result.}
  \label{fig_fit_result_scan}
\end{figure}

These results, apart from \Scpm, are the world's most precise measurements of the branching fractions and $CP$ violation parameters in \BK\ decays. To estimate the significance of $CP$ violation, we perform a 
two-dimensional likelihood scan in the \Acpm-\Scpm\ plane. This distribution is convolved with a two-dimensional Gaussian with means of zero and widths set to the relevant systematic uncertainty in \Acpm\ and \Scpm. 
The resulting distribution is then used to obtain contours in units of significance from which we find the first evidence for $CP$ violation in the \BoKs\ decay with 3.1 standard deviations, as shown in 
Fig.~\ref{fig_fit_result_scan}. 

As a test of the accuracy of the result, we perform a fit on the data set containing the first $535\times 10^{6}$ \BBbar\ pairs, which corresponds to the integrated luminosity used in the previous analysis. 
We obtain $\Acpm = -0.17 \pm 0.24$ and $\Scpm = +0.42 \pm 0.40$, which are in agreement with the previous Belle results shown in Table~\ref{tab_wks_prev}, considering the new tracking algorithm, the 37\% increase in 
detection efficiency with respect to that given in Ref.~\cite{wks_Belle} and the improved analysis strategy of including \Hel, which provides powerful discrimination between signal and background. Using a 
pseudoexperiment technique based on the fit result, we estimate the probability of a statistical fluctuation in the new data set causing the observed shift in the central value of \Scpm\ from our measurement with the 
first $535\times10^{6}$ \BBbar\ pairs to be 7\%.

\section{Systematic uncertainties}
\label{Systematic Uncertainties}
Systematic uncertainties from various sources are considered and estimated with both model-specific and -independent studies and cross-checks. All uncertainties are summarized in Table~\ref{tab_syst}. The systematic 
uncertainty due to the error on the total number of \BBbar\ pairs is calculated from the on- and off-resonance luminosity, taking into account efficiency and luminosity scaling corrections. The uncertainty arising 
from \piz\ reconstruction is evaluated by comparing data-MC differences of the yield ratio between $\eta\rightarrow\piz\piz\piz$ and $\eta\rightarrow\pip\pim\piz$. The uncertainties due to $K^{0}_{S}$ reconstruction 
and tracking efficiencies are calculated by comparing data-MC differences of the reconstruction efficiency of $D^{*}\rightarrow D^{0}[K^{-}\pip] \piz$. The uncertainty due to particle identification efficiency is 
determined using inclusive $D^{*+}\rightarrow D^{0}[K^{-}\pip]\pip$ decays, where the PID of each particle is unambiguously determined by the charge.

The vertices of \Brec\ and \Btag\ are constructed with an IP constraint smeared in the $x - y$ plane by 21 $\mu$m to account for the finite flight length of the $B$ meson. This systematic error is estimated by varying 
the amount of smearing by $\pm10$ $\mu$m. The track selection cut values on the tag side are varied by $\pm 10\%$, and the difference in the fit result is taken as a systematic uncertainty. The charged track 
parametrization errors are corrected by global scaling factors obtained from cosmic rays. The effect of these corrections is studied by looking at the difference in fit results with and without the corrected errors. 
The requirement of $|\Dt| < 70$ ps is varied by $\pm 30$ ps. The $B$ vertex quality selection criteria, $h < 50$, is varied by $^{+50}_{-25}$ and the $z$ vertex error requirements, $\sigma < 200 \; (500)$ $\mu$m for 
multi- (single-) track vertices is varied by $\pm 100$ $\mu$m. A \Dz\ bias can be caused by an unknown intrinsic misalignment within the SVD or relative misalignment 
between the SVD and CDC. This scenario is considered by generating MC with and without misalignment effects and taking the difference as a systematic error.

The fit model systematics in the signal PDF include the fixed physics parameters \taub\ and \Dmd, which are varied within their world-average uncertainties~\cite{PDG}. It also includes the \Dt\ resolution function 
parameters of $R^s_{\BzBzb}(\Dt)$ and $R^s_{\BpBm}(\Dt)$, as well as the flavor-tagging performance parameters $w$ and \Dw, which are varied within $\pm 1 \sigma$ of their experimental uncertainties determined from a 
control sample~\cite{jpsiks_Belle2,Tagging}. The fixed \BBbar\ background yields are also accounted for, where the nonpeaking background yields are varied within their MC errors, while the peaking background yields 
are varied taking into account the world-average uncertainties on their branching fractions. The parametric and nonparametric shapes describing the background are varied within their uncertainties. For nonparameteric 
shapes ({\it i.e.}, histograms), we vary the contents of the histogram bins by $\pm 1\sigma$. We vary the fractions of the Chebyshev and ARGUS 
components of the \De\ and \Mbc\ signal PDFs by their full amounts in order to estimate the uncertainty due to the presence of misreconstructed \piz\ and $K^{0}_{S}$ in the signal model. The systematic error due to 
the uncertainty of the relative yield of the misreconstructed signal component is estimated by varying its fraction by the full value estimated from MC simulation.

We study the uncertainties arising from $CP$ violation in the \BBbar\ background by introducing an artificial $CP$-violating component, which is set conservatively at 20\% of all neutral \BBbar\ events, and vary the 
$CP$ parameters maximally between $\Acpm=\pm1$ and $\Scpm=\pm1$. Half the fit bias obtained from pseudoexperiment MC studies is taken as an additional systematic uncertainty. A detector bias uncertainty is assigned to 
\Acpc, accounting for effects such as asymmetry in PID and tracking efficiencies, material effect using $D^+_s \to \phi [K^+K^-] \pi^+$ and $D^0 \to K^- \pi^+$ samples~\cite{detbias}. Finally, a large number of MC 
pseudoexperiments are generated, and an ensemble test is performed to obtain possible systematic biases from interference on the tag side arising between the CKM-favored $b \bar d \to (c \bar u d) \bar d$ and doubly 
CKM-suppressed $\bar b d \to (\bar u c \bar d) d$ amplitudes in the final states used for flavor tagging~\cite{tsi}.

\begin{table}[H]
  \small
  \renewcommand{\arraystretch}{1.5}%
  \centering
  \caption{Systematic uncertainties of the branching fractions and $CP$ asymmetries. The uncertainties on the $CP$ parameters are absolute, while those on the branching fractions are given as its
    percentage.}
  \begin{tabular}
    {@{\hspace{0.2cm}}l@{\hspace{0.2cm}}  @{\hspace{0.2cm}}c@{\hspace{0.2cm}}  @{\hspace{0.2cm}}c@{\hspace{0.2cm}}  @{\hspace{0.2cm}}c@{\hspace{0.2cm}}  @{\hspace{0.2cm}}c@{\hspace{0.2cm}}
      @{\hspace{0.2cm}}c@{\hspace{0.2cm}}}     
    \hline \hline 
    Category 	& $\delta{\cal B}(\omega K^{0})$ & $\delta\Acpm$ & $\delta\Scpm$ & $\delta{\cal B}(\omega K^{+})$ & $\delta\Acpc$  \\
    & $(\%)$ & $(10^{-2})$ & $(10^{-2})$ & $(\%)$ & $(10^{-2})$  \\
    \hline  	
    $N_{\BBbar}$ & 1.4 & N/A & N/A & 1.4 & N/A\\
    $\piz$ reconstruction & 4.0 & N/A & N/A & 4.0 & N/A\\
    $K^{0}_{S}$ reconstruction & 0.8 & N/A & N/A & N/A & N/A\\
    PID & 1.8 & N/A & N/A & 2.8 & N/A\\
    Tracking & 0.7 & N/A & N/A & 1.1 & N/A\\
    IP profile & 0.4 & 0.1 & 1.2 & 0.2 & N/A\\
    \Btag\ track selection & 0.5 & 0.2 & 0.3 & 0.2 & N/A\\
    Track helix error & 0.0 &  0.0 &  0.0 & 0.0 & N/A\\
    \Dt\ selection & 0.6 & 0.0 & 0.1 & 0.1 & N/A\\
    Vertex quality selection & 0.9 & 0.3 & 0.5 & 0.9 & N/A\\
    $\Delta z$ bias & N/A & 0.5 & 0.4 & N/A & N/A\\
    Misalignment & N/A & 0.4 & 0.2 & N/A & N/A\\
    Physics parameters & 0.0 & 0.1 & 0.1 & 0.0 & 0.0\\
    \Dt\ resolution function & 0.6 & 2.6 & 4.4 & 0.8 & 0.7\\
    Flavor tagging & 0.0 & 0.3 & 0.8 & 0.0 & N/A\\
    Misreconstruction & 0.9 & 0.1 & 0.3 & 0.7 & 0.1\\
    \BBbar\ background yields & 0.8 & 0.2 & 0.5 & 0.9 & 0.3\\
    Parametric shape & 1.8 & 0.5 & 1.5 & 1.0 & 0.5\\
    Nonparametric shape & 0.1 & 0.1 & 0.2 & 0.1 & 0.3\\
    Fit bias & 0.6 & 0.7 & 0.1 & 0.9 & 0.3\\
    Detector bias & N/A & N/A & N/A & N/A & 0.3\\
    Background $CP$ violation & N/A & 1.5 & 1.4 & N/A & 0.1\\
    Tag-side interference & N/A & 3.2 & 0.2 & N/A & N/A\\      
    \hline 	
    Total & 5.5& 4.6 & 5.2 & 5.6 & 1.0\\
    \hline \hline 	 	
  \end{tabular}
  \label{tab_syst}
\end{table} 

\clearpage
\section{Conclusion}
\label{Conclusion}
We report an improved measurement of the branching fraction and $CP$ violation parameters in \BK\ decays. The measurements are based on the full Belle data sample after reprocessing 
with a new tracking algorithm and with an optimized analysis performed with a simultaneous fit; they supersede those of the previous Belle analyses~\cite{wk_Belle,wks_Belle}. These are now the world's 
most precise measurements, apart from \Scpm, and the obtained values are mostly consistent with previous measurements from Belle and BaBar~\cite{wk_Belle,wks_Belle,wk_BaBar,wks_BaBar}, apart from a 
$3\sigma$ tension between the Belle and BaBar result for \Acpm. The results for the branching fractions, \Acpm\ and \Acpc, are in agreement with the predictions of the pQCD, QCDF, and SCET theories within 
one to two standard deviations. The value obtained for \Scpm\ is consistent with the prediction of the SM (see Table~\ref{tab_th_pred}) within one standard deviation, and the first evidence for $CP$ 
violation in \BoKs\ is found at the level of 3.1 standard deviations.
\section*{ACKNOWLEDGMENTS}

We thank the KEKB group for the excellent operation of the accelerator; the KEK cryogenics group for the efficient operation of the solenoid; and the KEK computer group,
the National Institute of Informatics, and the PNNL/EMSL computing group for valuable computing and SINET4 network support.  We acknowledge support from
the Ministry of Education, Culture, Sports, Science, and Technology (MEXT) of Japan, the Japan Society for the Promotion of Science (JSPS), and the Tau-Lepton Physics 
Research Center of Nagoya University; the Australian Research Council and the Australian Department of Industry, Innovation, Science and Research;
Austrian Science Fund under Grant No. P 22742-N16; the National Natural Science Foundation of China under Contract No.~10575109, No.~10775142, No.~10825524, No.~10875115, No.~10935008, and No.~11175187; 
the Ministry of Education, Youth and Sports of the Czech Republic under Contract No.~MSM0021620859; the Carl Zeiss Foundation, the Deutsche Forschungsgemeinschaft
and the VolkswagenStiftung; the Department of Science and Technology of India; the Istituto Nazionale di Fisica Nucleare of Italy; the WCU program of the Ministry of Education, Science and
Technology; National Research Foundation of Korea Grants No.~2011-0029457, No.~2012-0008143, No.~2012R1A1A2008330, and No.~2013R1A1A3007772; the BRL program under NRF Grant No. KRF-2011-0020333, 
No.~KRF-2011-0021196, the BK21 Plus program, and the GSDC of the Korea Institute of Science and Technology Information; the Polish Ministry of Science and Higher Education and the National Science Center;
the Ministry of Education and Science of the Russian Federation and the Russian Federal Agency for Atomic Energy; the Slovenian Research Agency;
the Basque Foundation for Science (IKERBASQUE) and the UPV/EHU under Program No. UFI 11/55; the Swiss National Science Foundation; the National Science Council
and the Ministry of Education of Taiwan; and the U.S.Department of Energy and the National Science Foundation. This work is supported by a Grant-in-Aid from MEXT for 
Science Research in a Priority Area (``New Development of Flavor Physics'') and from JSPS for Creative Scientific Research (``Evolution of Tau-lepton Physics'').

\end{document}